\documentclass[12pt]{iopart}

\newcommand{\apj}{{\it Astrophys. J.}}
\newcommand{\apjs}{{\it Astrophys. J. Supp.}}
\newcommand{\aap}{{\it Astron. \& Astrophys.}}
\newcommand{\aaps}{{\it Astron. \& Astrophys. Supp.}}
\newcommand{\nat}{{\it Nature}}
\newcommand{\mnras}{{\it MNRAS}}

\usepackage{graphicx}
\begin{document}

\title[Gamma ray astrophysics: the EGRET results]{Gamma ray astrophysics: the EGRET results}

\author{D J Thompson}

\address{Astrophysics Science Division, NASA Goddard Space Flight Center, Greenbelt, MD 20771 USA }
\ead{David.J.Thompson@nasa.gov}

\begin{abstract}
Cosmic gamma rays provide insight into some of the most dynamic processes in the Universe.  At the dawn of a new generation of gamma-ray telescopes, this review summarizes results from the Energetic Gamma Ray Experiment Telescope (EGRET) on the Compton Gamma Ray Observatory, the principal predecessor mission studying high-energy photons in the 100 MeV energy range.  EGRET viewed a gamma-ray sky dominated by prominent emission from the Milky Way, but featuring an array of other sources, including quasars, pulsars, gamma-ray bursts, and many sources that remain unidentified.  A central feature of the EGRET results was the high degree of variability seen in many gamma-ray sources, indicative of the powerful forces at work in objects visible to gamma-ray telescopes.
\end{abstract}

\maketitle

\section{Introduction: gamma rays from the Universe}
For most of history, humans have learned about the cosmos by viewing the light that our eyes can detect.  Only in the Twentieth Century did it become clear that vast amounts of information arrive in different channels, most prominently the invisible forms of electromagnetic radiation.  This information revolution started with radio, which can reach the Earth's surface, and then exploded with the space age discoveries that essentially every part of the spectrum carries unique information about conditions and processes in distant parts of the Universe. Most of these forms of radiation are blocked by the Earth's atmosphere and therefore require space observatories or some form of indirect detection. 

Gamma rays represent the high-energy end of the electromagnetic spectrum, comprising photons with the highest frequencies or shortest wavelength.  Because gamma rays are so energetic, they are usually defined by their energies, which are typically higher than 100 kilo electron Volts (keV), although the line between X-rays and gamma rays is not a sharp one.  Often X-rays are considered to be those photons produced by atomic or thermal processes, while gamma rays are those involving nuclear or nonthermal processes. Above 10$^6$ eV (1 MeV), and with no upper limit to energy, all photons are gamma rays.  

As might be imagined, gamma rays are produced by energetic phenomena.  In fact, only the lowest-energy gamma rays, those associated with radioactivity, have natural sources on Earth.  Cosmic sources of gamma rays extend to vastly higher energies, reflecting extreme physical conditions and powerful collisions.  Astrophysical settings for gamma-ray production include supernovae, pulsars, and quasars, as well as the interstellar and intergalactic medium.  Although gamma rays are absorbed in the atmosphere, the Universe is largely transparent to these high-energy photons out to high redshift.  Gamma rays thus provide a valuable probe of the largest energy transfers throughout much of the Universe.  Such phenomena can be expected to be important in understanding the forces of change on the largest scale. 

The present review is focused on one important segment of gamma-ray astrophysics: the results from the Energetic Gamma Ray Experiment Telescope (EGRET) that flew on the Compton Gamma Ray Observatory (CGRO) during its 1991-2000 life.  EGRET provided the first detailed all-sky observations of high-energy gamma rays (with typical energies in the 100 MeV to 1000 MeV range). With the advent of two successors to EGRET, the Italian Astro-rivelatore Gamma a Immagini Leggero (AGILE) mission and the international Fermi Gamma-ray Telescope (formerly the Gamma-ray Large Area Space Telescope, GLAST) mission, the time seems appropriate to examine the observational foundation of these two current missions. 

The outline of this review is as follows:

1.  Introduction
 
2. EGRET in the context of other gamma-ray telescopes
      
\indent    \indent   Historical
       
\indent    \indent    CGRO
       
\indent  \indent     EGRET:  Detecting high-energy gamma rays
       
3. An overview of the gamma-ray sky

4. Galactic diffuse

5. Gamma-ray sources: the third EGRET catalogue

6. Galactic sources 

\indent   \indent  Pulsars
     
\indent  \indent   Binary Sources

\indent  \indent   Other Sources

7. Extragalactic sources

 \indent   \indent  Blazars
     
\indent   \indent  Other galaxies
     
\indent  \indent    Diffuse extragalactic radiation
     
\indent  \indent   Gamma-ray bursts
     
8. Local sources 

\indent   \indent  The Moon
     
\indent  \indent    Solar flares
     
\indent  \indent   Primordial black holes

9. Open questions for AGILE and Fermi

\section{EGRET in the context of other gamma-ray telescopes}

\subsection{Before the Compton Observatory}

The potential for studying the sky with gamma rays was identified fifty years ago (e.g.  Morrison 1958).  A decade later the first definitive detections of gamma rays from space came with the OSO-3 discovery of gamma radiation from the plane of our Galaxy (Clark \etal1968).  In following years, a number of fairly small balloon-borne and satellite missions began to reveal aspects of the gamma-ray sky.  Some examples:

\begin{itemize}
\item Browning, Ramsden and Wright (1971) found pulsed high-energy gamma radiation from the Crab Pulsar, using a gamma-ray telescope carried on a balloon.
\item The U.S. Small Astronomy Satellite 2 (SAS-2, Fichtel \etal 1975) showed that high-energy gamma rays help trace the structure of our Galaxy, the Milky Way (Hartman \etal 1979).  It also discovered a second gamma-ray pulsar, the Vela Pulsar (Thompson \etal 1975) and the first unidentified gamma-ray source, $\gamma$195+5 (Kniffen \etal 1975), which later was called Geminga (Bignami \etal 1983).
\item The European COS-B satellite (Bignami \etal 1975) produced the first catalogues of high-energy gamma-ray sources, most of which were not identified with objects seen at other wavelengths (Hermsen \etal 1977, Swanenburg \etal 1981).  In addition, it found the first extragalactic gamma-ray source, the nearby quasar 3C273 (Swanenburg \etal 1978) and made the first gamma-ray observations of molecular clouds as spatially-extended sources (Caraveo \etal 1980).
\item VELA satellites operated by the U.S. military discovered short-duration cosmic gamma-ray bursts, a completely new phenomenon (Klebesadel \etal 1973).
\item The Third High Energy Astrophysical Observatory (HEAO-3) carried a low-energy gamma-ray telescope with high spectral resolution  (Mahoney \etal 1980) that detected the 0.5 MeV positron-electron annihilation line coming from the Galactic Center region (Riegler \etal 1981), confirming previous reports from balloon instruments (e.g. Leventhal \etal 1978).
\end{itemize}

\begin{figure}
\centering
\includegraphics{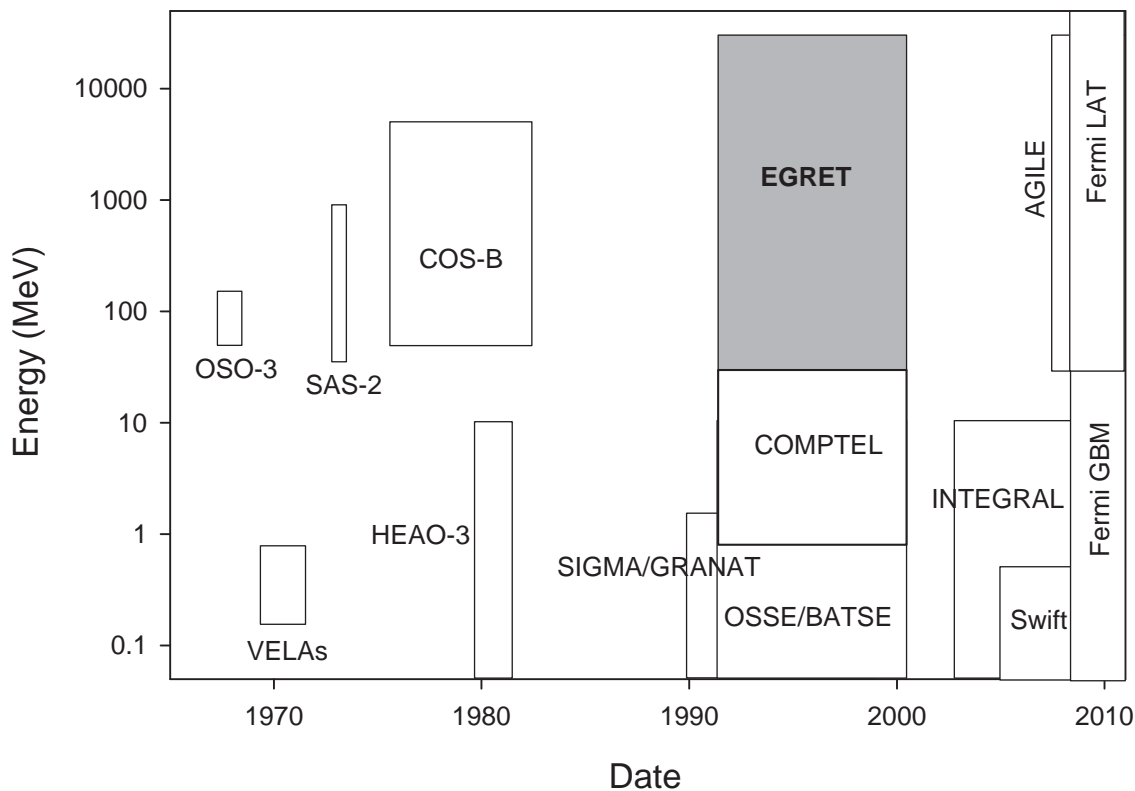}
\caption{Some space gamma-ray missions, showing energy coverage and the time frame of the mission.}
\label{fig:figure1}
\end{figure}

During this same period, a parallel branch of gamma-ray astrophysics using ground-based detectors was developing.  At gamma-ray energies above about 10$^{11}$ eV (100 GeV), cosmic photons are too scarce to be detected by satellite detectors.  The Earth's atmosphere itself can be used, however, as a detector for these very high energy (VHE) photons.  When such photons collide with the material at the top of the atmosphere, they produce showers of particles moving faster than the local speed of light, thereby emitting Cherenkov radiation in the optical and ultraviolet.  The flashes of light from these interactions can be detected with Atmospheric Cherenkov Telescopes (ACTs) on the ground, providing an indirect way to conduct gamma-ray astrophysics. A milestone in VHE gamma-ray astrophysics was reached in 1989 with the high-confidence detection of the Crab Nebula (but not the pulsar) with the Whipple Observatory ACT (Weekes \etal 1989).  Although VHE studies are beyond the scope of the present review, this rapidly-developing field is highly complementary to space-based gamma-ray studies.  Weekes (2003) provides a broad review of gamma-ray astrophysics, with emphasis on the VHE field, while Chadwick \etal (2008) and Aharonian \etal (2008) present recent summaries of results.

Each of these programs provided a glimpse of the gamma-ray Universe, confirming the potential of this field for helping understand high-energy cosmic phenomena.  None of them gave a comprehensive picture.  Some viewed only portions of the sky.  Some studied only a limited portion of the huge energy range falling under the gamma-ray banner.  Most were operated for only limited durations.  Figure 1 shows schematically the time frames and energy ranges of many of these space-based gamma-ray telescopes.  NASA's Great Observatories program offered the opportunity to make a major advance across much of the gamma ray portion of the spectrum.

\subsection{The Compton Gamma Ray Observatory}

The Compton Gamma Ray Observatory (CGRO), shown in Figure 2, was the second of NASA's Great Observatories, following the Hubble Space Telescope.  Although the original plan was to have all four operating simultaneously, circumstances delayed the launches of the Chandra X-ray Observatory (originally called AXAF) and the Spitzer Space Telescope (Infrared, originally called SIRTF) until the Compton Observatory mission was essentially complete. CGRO itself was launched on the Space Shuttle Atlantis on 5 April 1991 and operated successfully until it was de-orbited on 4 June 2000.  

\begin{figure}
\centering
\includegraphics [width = 5.5 in.] {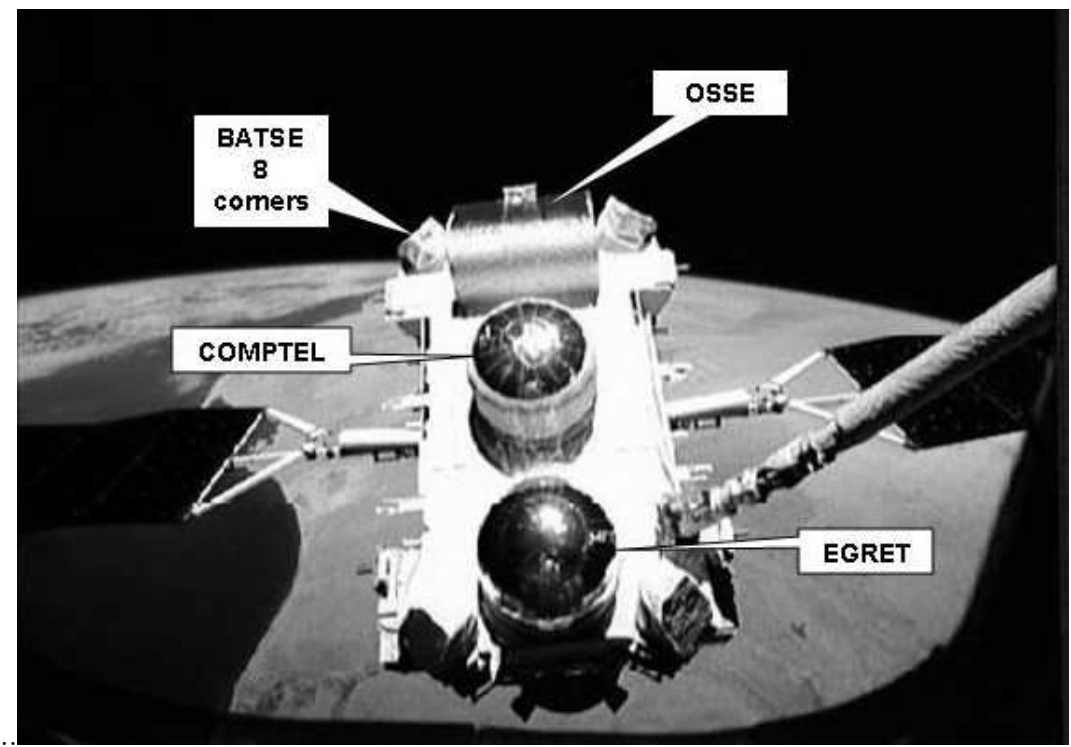}
\caption{The Compton Gamma Ray Observatory just before its release by the Shuttle in April. 1991.}
\label{fig:figure2}
\end{figure}

CGRO carried four gamma-ray telescopes, each with its own energy range, detection technique, and scientific goals.  Together these four instruments covered energies from less than 15 keV to more than 30 GeV, over six orders of magnitude in the electromagnetic spectrum. The three lower-energy telescopes were:

\begin{itemize}
\item Burst and Transient Source Explorer (BATSE, Principal Investigator, G. Fishman, NASA Marshall Space Flight Center). BATSE was the smallest of the CGRO instruments, consisting of one module located on each corner of the spacecraft.  Each BATSE unit included a large flat NaI(Tl) scintillator pointed with its face away from the center of the observatory and a  smaller thicker scintillator for spectral measurements, combining to cover an energy range from 15 keV to over 1 MeV.  Important results from BATSE included the mapping of over 2700 gamma-ray bursts, showing an isotropic distribution on the sky. A summary of BATSE results is given by Fishman (1995). 
\item Oriented Scintillation Spectrometer Experiment (OSSE, Principal Investigator J. Kurfess, Naval Research Laboratory).  OSSE used four large, collimated scintillator detectors to study low-energy gamma-rays, 60 keV - 10 MeV.  OSSE mapped the 0.5 MeV line from positron annihilation and provided detailed measurements of many hard X-ray/soft-gamma-ray sources.  Kurfess (1996) summarized many of the important results from OSSE.
\item Imaging Compton Telescope (COMPTEL, Principal Investigator V. Sch\"onfelder, Max Planck Institute for Extraterrestrial Physics). COMPTEL detected medium-energy gamma rays using a Compton scattering technique, effective between 0.8 MeV and 30 MeV.  Among its results, COMPTEL mapped the distribution of radioactive Aluminum-26 in the Galaxy, showing the locations of newly formed material. The summary by Sch\"onfelder \etal (1996) describes many of the COMPTEL results.
\end{itemize}

This brief summary covers only a few of the many important results from these CGRO instruments.  Sch\"onfelder (2001) reviewed the entire field of gamma-ray astrophysics, with particular emphasis on this energy range.

\subsection{EGRET: Detecting High-Energy Gamma Rays}

In the energy range above 10 MeV, the principal interaction process for gamma rays is pair production, a direct example of the Einstein equation E = mc$^2$.  The photon energy is converted into an electron and its antiparticle, a positron. This process can take place in the field of an atomic nucleus or in a strong magnetic field, but not in free space, in order to conserve energy and momentum.  These high-energy gamma rays cannot be reflected or refracted; a gamma-ray telescope actually detects the electron and positron. 

\begin{figure}
\centering
\includegraphics [width = 6.0 in.] {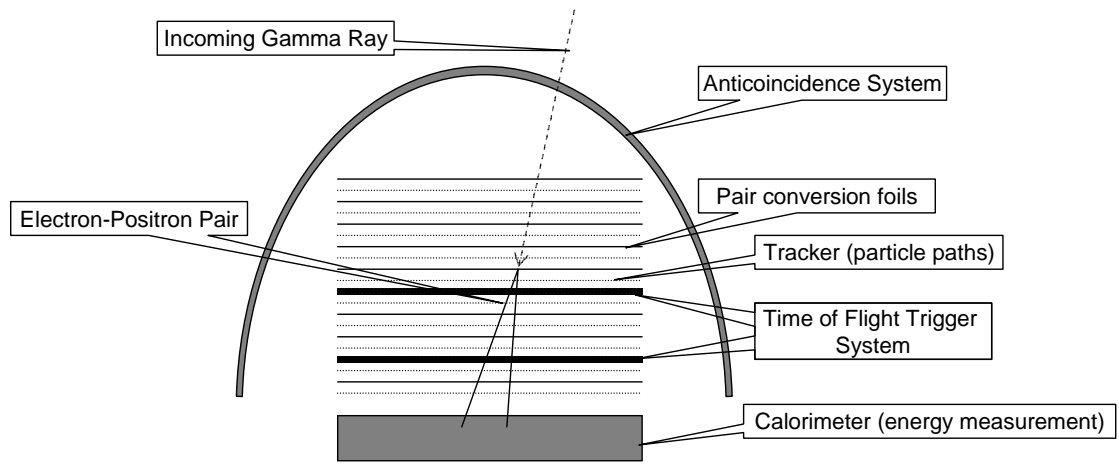}
\caption{Schematic diagram of a pair production telescope.}
\label{fig:figure3}
\end{figure}

The Energetic Gamma Ray Experiment Telescope (EGRET) was the high-energy instrument on the Compton Observatory, covering the energy range 20 MeV to 30 GeV (Hughes \etal 1980; Fichtel \etal 1983).  The Co-Principal Investigators represented the three major contributors to the EGRET hardware:  C. Fichtel, NASA Goddard Space Flight Center (GSFC); R. Hofstadter, Stanford University (SU); and K. Pinkau, Max Planck Institute for Extraterrestrial Physics (MPE).  The operational concept of EGRET, similar in most respects to the designs of other high-energy gamma-ray telescopes, is shown in Figure 3.  The two key challenges for any such telescope are: (1) identify the gamma-ray interaction in the presence of a huge background of charged particles (cosmic rays, solar particles, and trapped radiation); and (2) measure the gamma-ray arrival time, arrival direction, and energy.

The process works as follows:
\begin{enumerate}
\item A gamma ray enters EGRET. It first passes through the Anticoincidence System without producing a signal.
\item The gamma ray interacts in one of 28 thin tantalum sheets. This interaction converts the gamma ray into an electron and a positron via pair production.
\item The spark chamber Tracker records the paths of the electron and positron, allowing EGRET to see the pair interaction and to determine the arrival direction of the gamma ray.
\item The electron and positron pass through two scintillator detectors operated in a time-of-flight (TOF) configuration.  The TOF signal confirms the direction of the particles and triggers the readout of the spark chambers.
\item The electron and positron enter the Calorimeter, producing an electromagnetic shower, which measures the energies of the particles and therefore the energy of the original gamma ray.
\item Unwanted cosmic-ray particles produce signals in the Anticoincidence System, which tell the electronics not to trigger the spark chamber. The Anticoincidence System rejects nearly all unwanted signals produced by cosmic rays that enter EGRET.
\end{enumerate}

\begin{figure}
\centering
\includegraphics [width = 6.0 in.,clip=true] {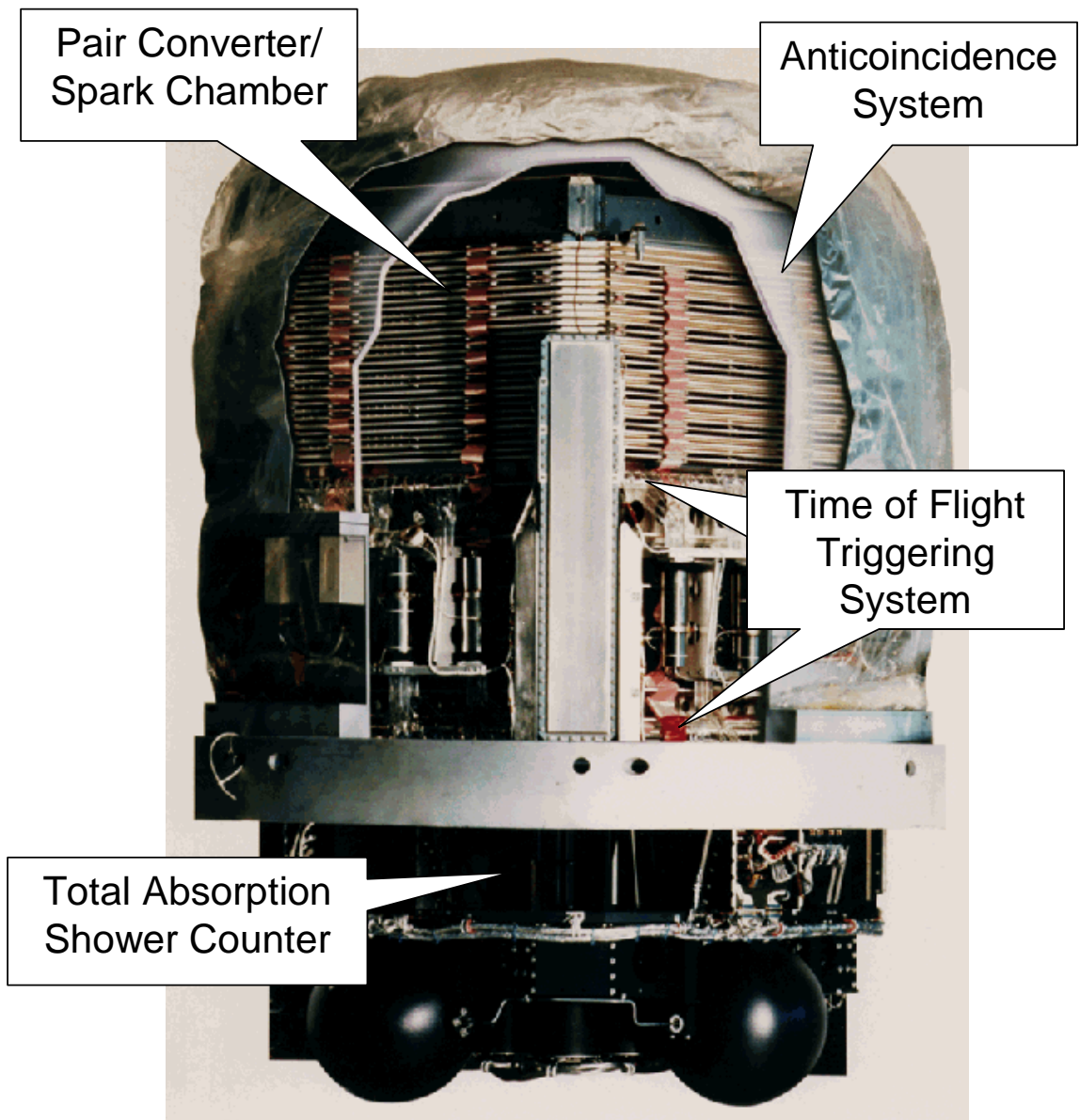}
\caption{Composite photo showing a cutaway view of EGRET.  The major subsystems are identified. }
\label{fig:figure4}
\end{figure}

Figure 4 shows EGRET as it appeared before integration onto the Compton Gamma Ray Observatory.  Here is some information about the key subsystems:

\begin{itemize}
\item The Anticoincidence System consists of a single dome of plastic scintillator, read out by 24 photomultiplier tubes mounted around the bottom.  This subsystem was built by the MPE group. 
\item The Tracker is made of 36 wire grid spark chambers, interleaved with the converter plates.  The active area of the spark chambers is 81 cm x 81 cm.  This subsystem was the work of the group at GSFC. 
\item The TOF trigger system, also from GSFC, has two four by four arrays of plastic scintillator tiles, each read out by a single photomultiplier tube.  
\item The Calorimeter, called the Total Absorption Shower Counter (TASC), was made of 36 NaI crystals bonded together and read out by 16  photomultiplier tubes.  The SU group  built the TASC.
\end{itemize}

The primary EGRET calibration was carried out at the Stanford Linear Accelerator Center (SLAC), using a gamma-ray beam varying in energy from 15 MeV to 10 GeV.  The beam scanned the active area of EGRET at a variety of angles out to 40$^{\circ}$ from the instrument axis. The calibration thus covered essentially the entire phase space to be observed in orbit (Thompson \etal 1993). Analysis of  both the calibration and flight  data concentrated on recognizing and measuring the individual pair-production events.  An initial selection of events was made in software, removing many triggers that did not produce tracks consistent with being electron-positron pairs. Events the software could not resolve were reviewed by data analysts and scientists, leading to a final data set almost entirely free of charged particle contamination. 

The basic properties of EGRET are shown in Table 1.  The single-photon angular resolution, or point spread function (PSF) is energy-dependent.  The angle $\theta$ in degrees containing 67\% of the gamma rays from a delta function source with energy E in MeV  is given by:  $\theta$ = 5$^\circ$.85 E$^{-0.534}$.

\begin{table}
\caption{\label{label}Performance Characteristics of EGRET.}
\begin{indented}
\item[]\begin{tabular}{@{}ll}
\br
Property&Value\\
\mr
Energy Range&20 MeV - $>$ 10 GeV\\
Peak Effective Area&1500 cm$^2$ at 500 MeV \\
Energy Resolution&15\% FWHM\\
Off-axis effective area&25\% of peak at 30$^{\circ}$\\
Timing accuracy&$<$ 100 $\mu$sec absolute\\
\br
\end{tabular}
\end{indented}
\end{table}

Observations of the Compton Gamma Ray Observatory ranged in duration from a few days to a few weeks, reflecting the paucity of cosmic gamma-ray photons.  A listing of the observations, along with other information about CGRO, can be found at the CGRO Science Support Center Web site, {\it http://cossc.gsfc.nasa.gov/docs/cgro/index.html}.

Because the EGRET spark chambers were gas detectors, their performance deteriorated with time due to gas aging.  The system carried enough replacement gas for four complete refills of the chamber, all of which were used.  The performance of the instrument was recalibrated in flight, using constant sources such as the diffuse Galactic emission and bright pulsars as reference sources.  The preliminary in-flight calibration was described by Esposito \etal 1999, and the final performance analysis was described by Bertsch \etal 2001.

\section{An overview of the high-energy gamma-ray sky}

In its nine-year lifetime, EGRET detected over 1,500,000 celestial gamma rays. One photon at a time, EGRET built up a picture of the entire high-energy gamma-ray sky.  Figure 5 shows the summed map above 100 MeV, in Galactic coordinates. The Milky Way runs horizontally across the center of the figure, and the Galactic center lies at the center of the map. 

\begin{figure}
\centering
\includegraphics [height = 6.5 in., angle = 90] {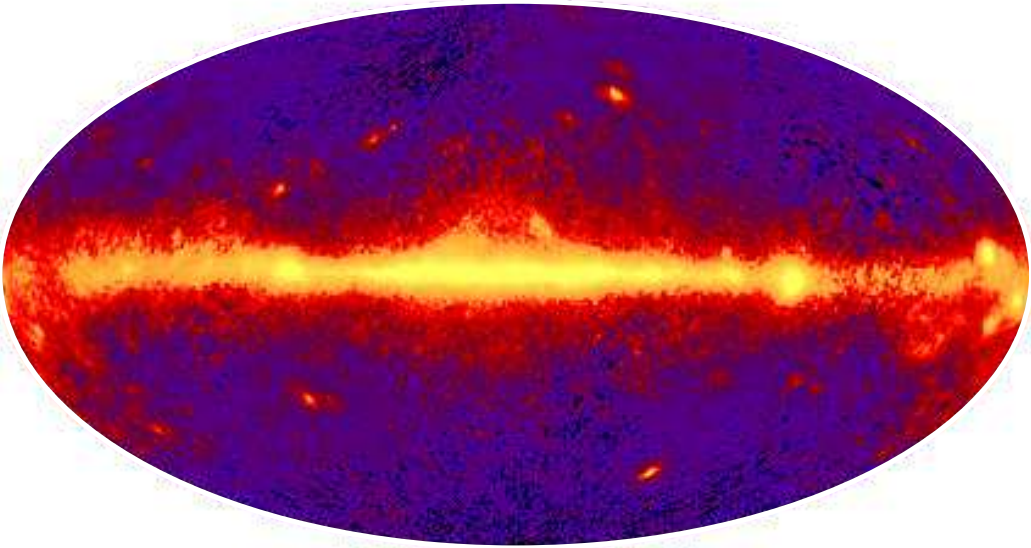}
\caption{The sky seen with EGRET, shown in Galactic coordinates.  In this false color image, the Galactic Center lies in the middle of the image.}
\label{fig:figure5}
\end{figure}

This image provides a striking contrast to the view of the sky at visible wavelengths.  Some of the key features that will be discussed in following sections are:
\begin{itemize}
\item The Milky Way is extremely bright, particularly toward the inner part of the Galaxy. 
\item The brightest persistent sources are pulsars.
\item Many of the bright sources away from the Galactic plane are highly variable types of Active Galactic Nuclei in the blazar class. 
\item The Moon is visible and outshines the Sun most of the time. 
\item Many of the sources are not identified with known objects. 
\end{itemize}

\section{Galactic diffuse gamma-ray emission}

The high-energy gamma-ray sky is dominated by the bright Galactic ridge.  This component was the first one predicted and detected in  gamma-ray astronomy, because our galaxy is known to be filled with high-energy particles, magnetic fields, photon fields, and interstellar gas.  The spatial distribution of the gamma radiation traces galactic structure as determined from radio and other measurements.  These basic features were known in the SAS-2 and COS-B era, along with the realization that the cosmic ray flux cannot be uniform throughout the Galaxy and still be consistent with these measurements.  The EGRET data provided detailed information, but also introduced a puzzle that remains unresolved. 

 The physical processes that produce EGRET-energy gamma rays are familiar ones to the particle physics community:

\begin{itemize}
\item Inelastic collisions of cosmic ray particles with the interstellar gas (mostly hydrogen) produce secondary particles, particularly charged and neutral $\pi$ mesons.  The neutral pions decay almost immediately into two gamma rays.  In the center of mass reference frame, each gamma ray has half the energy of the $\pi^{\circ}$, or about 67 MeV.  Because the cosmic rays typically have a broad range of high energies, the energy spectrum of the gamma rays  resulting from nucleon-nucleon collisions is spread out into a broad peak rather than a narrow line.
\item Cosmic ray electrons colliding with photons can boost the photon energies into the gamma-ray band by inverse Compton scattering.  The principal targets are the optical and infrared photons found throughout the Galaxy. 
\item Another cosmic-ray electron process involves collisions with the interstellar gas, producing gamma rays through bremsstrahlung. 
\item Both nucleon and electron cosmic rays can in principle produce gamma rays through interactions with magnetic fields by synchrotron radiation.  In practice, the fluxes expected from synchrotron radiation with known particles and magnetic fields are small compared the the other sources. 
\end{itemize}
 
 \begin{figure}
\centering
\includegraphics [height = 6.5 in.] {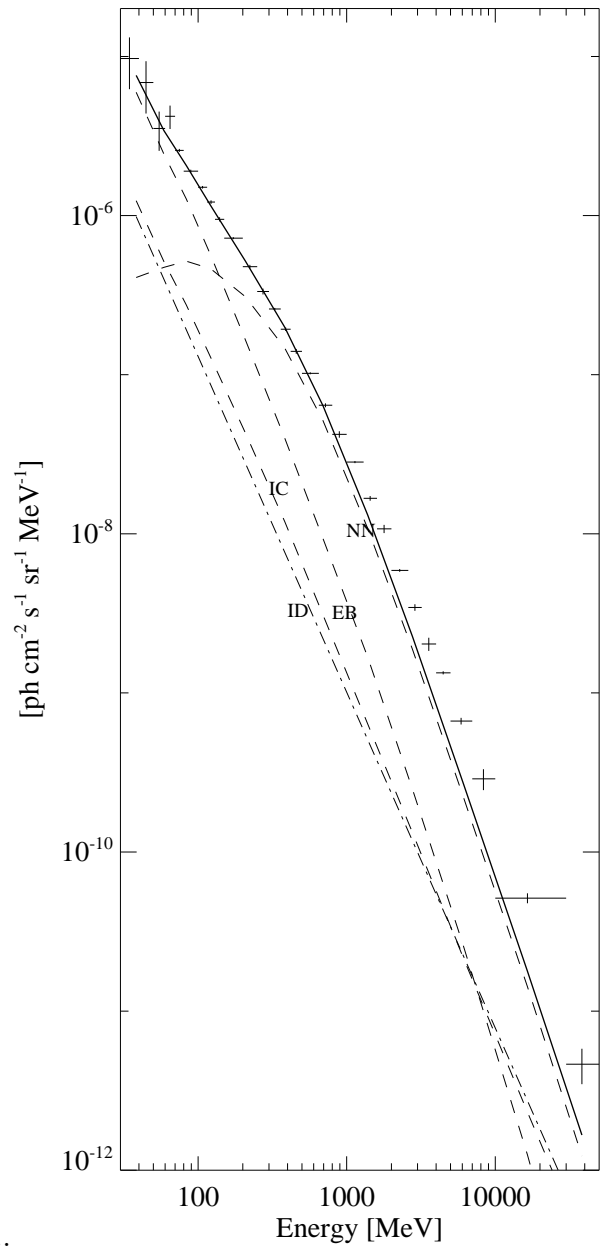}
\caption{Diffuse Galactic spectrum (Hunter \etal 1997).  Calculated components are NN: nucleon-nucleon interactions ( $\pi^{\circ}$-decay); EB: electron bremsstrahlung; IC: Inverse Compton; ID: Isotropic diffuse (see section 7.3).}
\label{fig:figure6}
\end{figure}
 
The EGRET energy spectrum of the gamma radiation from the Galactic Center region is shown in Figure 6, along with the calculated source components (Hunter \etal 1997).  Below 100 MeV, electron bremsstrahlung is the principal component, while at higher energies the nucleon-nucleon $\pi^{\circ}$-decay source is the most important.  The expected ``bump'' compared to a power law spectrum is clearly visible. 

 Building on the work of Bertsch \etal (1993), Hunter \etal (1997) carried out detailed modeling of the Galactic radiation.   Measured cosmic-ray intensities were combined with a three-dimensional model of the photon and gas distribution, using 21 cm radio measurements to trace neutral hydrogen (HI) and carbon monoxide (CO) transition radio measurements as a tracer of molecular hydrogen (H$_2$).  Coupling between the matter and cosmic ray densities was assumed, based on arguments of dynamic balance between matter, magnetic fields, and cosmic rays.  This model reproduced most features of the observed gamma radiation. 
 
 Visible in Figure 6 is one unexpected feature of the EGRET observations:  the flux above 1 GeV exceeds the model prediction by a significant amount (well beyond any known measurement uncertainties). This discrepancy has become known as the ``GeV excess.''
  
 An alternative modeling approach, called GALPROP, has been developed by Strong, Moskalenko, and Reimer (2000, 2004b).  This model emphasizes cosmic-ray propagation calculations and a larger inverse Compton contribution to the gamma radiation. Although the basic version of GALPROP does not reproduce the GeV excess, these authors have shown that plausible assumptions about cosmic ray densities in the Galaxy higher than the local values can reproduce the EGRET data. 
 
 De Boer (2005) has invoked a new component of the gamma radiation coming from annihilation of supersymmetric dark matter as the source of the GeV excess. Bergstr\"om  \etal (2006) conclude, however, that this dark matter model is inconsistent with measurements of antiprotons. Stecker, Hunter, and Kniffen (2008) argue in favor of a miscalibration of the EGRET detector at high energies as an explanation. No consensus exists.  It is perhaps ironic that the one component of the gamma-ray sky that was thought to be understood has left a mystery at the end of the EGRET mission. 
 
On a more local scale, Hunter \etal (1994), Digel, Hunter and Mukherjee (1995), and Digel \etal (1996) conducted gamma-ray studies of the nearby Ophiuchus, Orion, and Cepheus regions.  By comparing the EGRET maps to maps of CO clouds, they were able to trace the ratio of molecular hydrogen column density to integrated CO intensity.   The gamma-ray intensities were found consistent with models based on the local flux measurements of cosmic rays. 
 
Another issue with the diffuse radiation is that the matter content of the Galaxy is not necessarily completely measured.  Infrared data suggest the presence of unseen gas clouds that could require an additional component in the diffuse Galactic gamma-ray model (Grenier \etal 2005).

\section{Gamma-ray sources: the third EGRET catalogue}

Individual gamma-ray sources appear as excesses above the modeled diffuse emission.  The EGRET analysis process used a maximum likelihood method to compare probabilities of fitting a given region of the sky with and without a source (Mattox \etal 1996).  The most complete analysis of the sky by the EGRET team was the third EGRET catalogue (3EG: Hartman \etal 1999).  Starting from the Hunter \etal (1997) diffuse model, the 3EG analysis examined each viewing period plus combinations of viewing periods, from the start of the mission up through the end of 1995, using multiple energy ranges.  Because EGRET was operated only intermittently after this time, the total exposure added by later phases of the mission contributed little to the overall mapping of the sky. 

Figure 7 summarizes the 3EG results on gamma-ray sources, with the source locations shown in Galactic coordinates.  In this figure, the symbol size indicates the peak source brightness.  The gamma-ray sky is highly variable, so not all sources were seen at all times.  The census of the 271 3EG gamma-ray sources was:
\begin{itemize}
\item 94 sources show a probable or possible association with the class of Active Galactic Nuclei known as blazars. 
\item Five pulsars appear in the catalogue.
\item The Large Magellanic Cloud was detected as an extended gamma-ray source.
\item One solar flare was bright enough to be seen in the source analysis
\item 170 sources, well over half the total, had no identification with known astrophysical objects.
\end{itemize}

Following the publication of the 3EG catalogue, extensive efforts were made to identify individual sources or source populations.  A few of the 3EG sources appear to be artifacts of the analysis (Thompson \etal 2001). The following sections describe the detected source classes and efforts at identification, as well as some sources that did not appear in the catalogue. 

Recently, Cassandjian and Grenier (2008) have developed a new catalogue of EGRET sources, based on a new model of the diffuse emission (Grenier \etal 2005).  This catalogue, which contains only 188 sources, incorporates many of the 3EG sources into the diffuse radiation as gas concentrations, particularly at intermediate Galactic latitudes.  It does, however, remove a number of identified sources, some of which showed evidence of time variability.  Changing the diffuse model would not be expected to eliminate time-variable sources. This catalogue should probably be considered as an alternative analysis rather than a replacement for the 3EG catalogue.  Because it is so new, it has not received the same level of study as the 3EG catalogue. 

 \begin{figure}
\centering
\includegraphics [height = 6.5 in., angle =90] {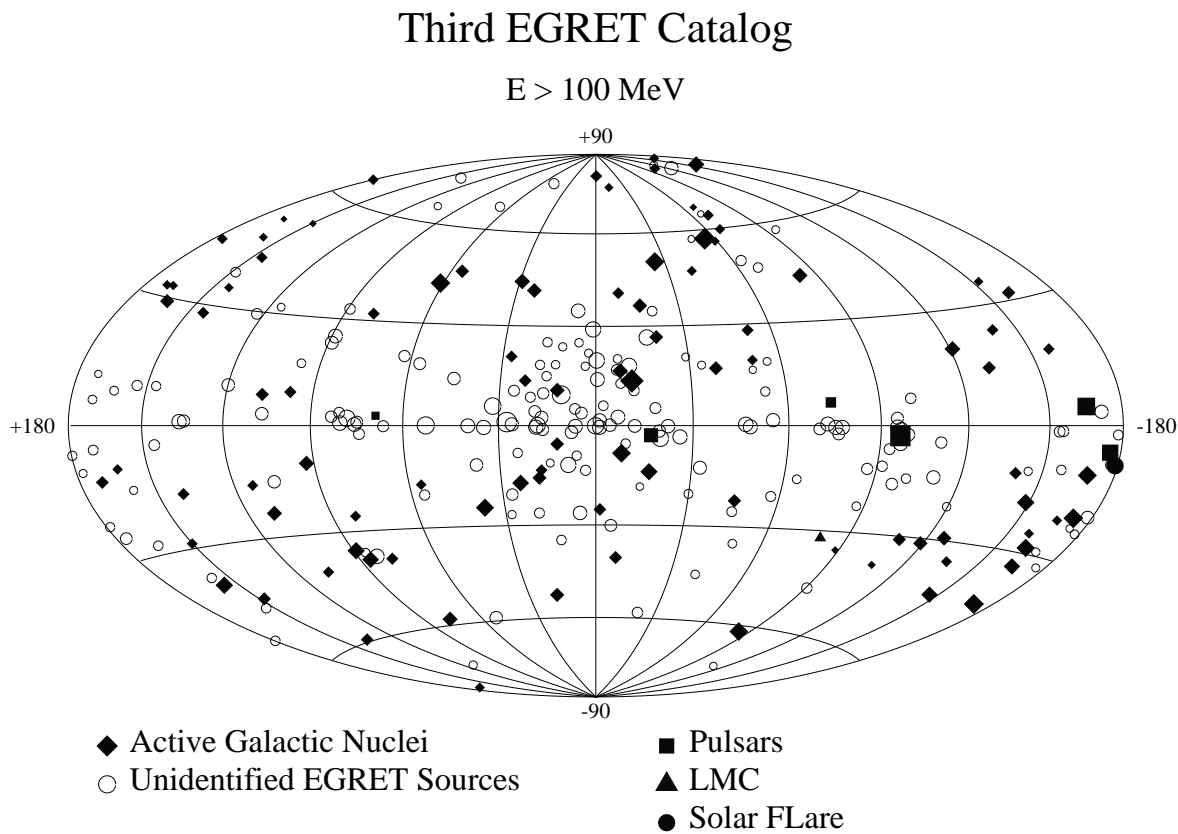}
\caption{Map of source locations for the third EGRET catalogue (Hartman \etal 1999), shown in Galactic coordinates.}
\label{fig:figure7}
\end{figure}

\section{Galactic gamma-ray sources}

Particularly along the Galactic Plane, the modeling of the diffuse gamma radiation strongly affects the calculated properties, or even the existence, of sources.  For this reason, the third EGRET catalogue  adopted different confidence levels for including sources as detections.  Within 10 degrees of the Galactic plane, a statistical significance of 5 $\sigma$ was required; at higher latitudes the requirement was 4 $\sigma$.  Even with this more stringent requirement, Figure 7 shows a clear population of sources concentrated along the plane.

\subsection{Pulsars}

The first high-energy gamma-ray source class was rotation-powered pulsars, starting with the Crab and Vela, seen by SAS-2 and COS-B.  EGRET expanded the number of gamma-ray pulsars to at least 6, with several other good candidates.  A summary of the EGRET results on pulsars is given by Thompson (2004). These rapidly-rotating neutron stars, originally seen in the radio (Hewish \etal 1968), have strong magnetic, electric, and gravitational fields.  Particles accelerated to high energies in the magnetospheres of pulsars can interact near the pulsar to produce gamma rays through curvature radiation, synchrotron radiation, or inverse Compton scattering.  

The telescopes on the Compton Gamma Ray Observatory identified seven or more gamma-ray pulsars, some with very high confidence and others with less certainty.    Figure 8 shows the light curves from the seven highest-confidence gamma-ray pulsars in five energy bands: radio, optical, soft X-ray ($<$1 keV), hard X-ray/soft gamma ray ($\sim$10 keV - 1 MeV), and hard gamma ray (above 100 MeV).  Based on the detection of pulsations, all seven of these are positive detections in the gamma-ray band. 

\begin{figure}
\centering
\includegraphics [height = 6.0 in.] {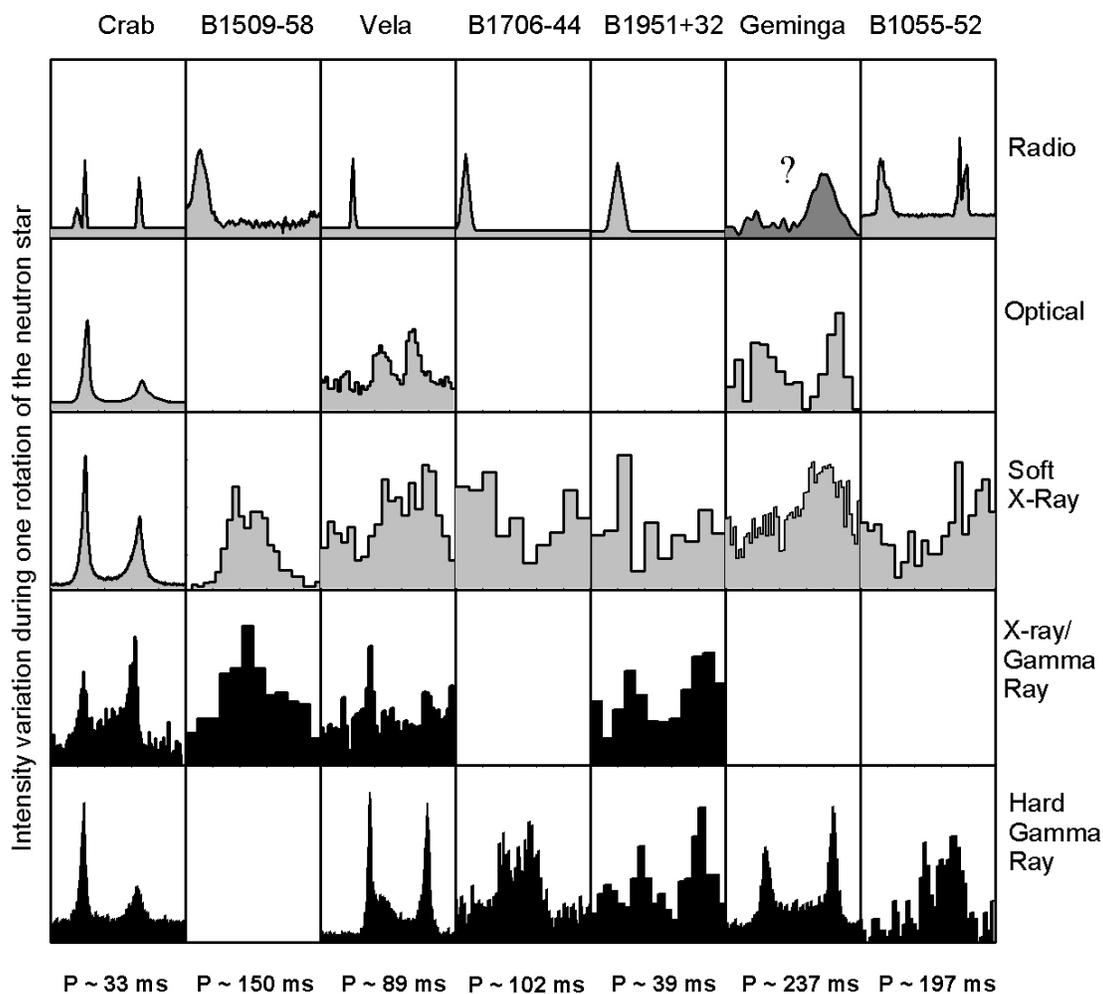}
\caption{Light curves of seven gamma-ray pulsars in five energy bands. Each panel shows one full rotation of the neutron star.  Adapted from Thompson (2004).}
\label{fig:figure8}
\end{figure}

Some important features of these pulsar light curves are:
\begin{itemize}
\item 
 They are not the same at all wavelengths.  Some combination of the geometry and the emission mechanism is energy-dependent.  In soft X-rays, for example, the emission in some cases appears to be thermal, probably from the surface of the neutron star;  thermal emission is not the origin of radio or gamma radiation.

\item Not all seven are seen at the highest energies.  PSR B1509$-$58 is seen up to 10 MeV by COMPTEL (Kuiper et al. 1999), but not at higher energies by EGRET.  

\item The six seen by EGRET all have a common feature - they show a double peak in their light curves.  Because these high-energy gamma rays are associated with energetic particles, it seems likely that the particle acceleration and interactions are taking place along a large hollow cone or other surface.  Models in which emission comes from both magnetic poles of the neutron star appear less probable in light of the prevalence of double pulses. 

\end{itemize}

In addition to the six high-confidence pulsar detections above 100 MeV, three additional radio pulsars may have been seen by EGRET:  PSR B1046$-$58, PSR B0656+14, and PSR J0218+4232,  the only millisecond pulsar with evidence of gamma-ray emission (Kuiper et al 2000, 2002). 

\begin{figure}
\centering
\includegraphics [height = 8.5 in.] {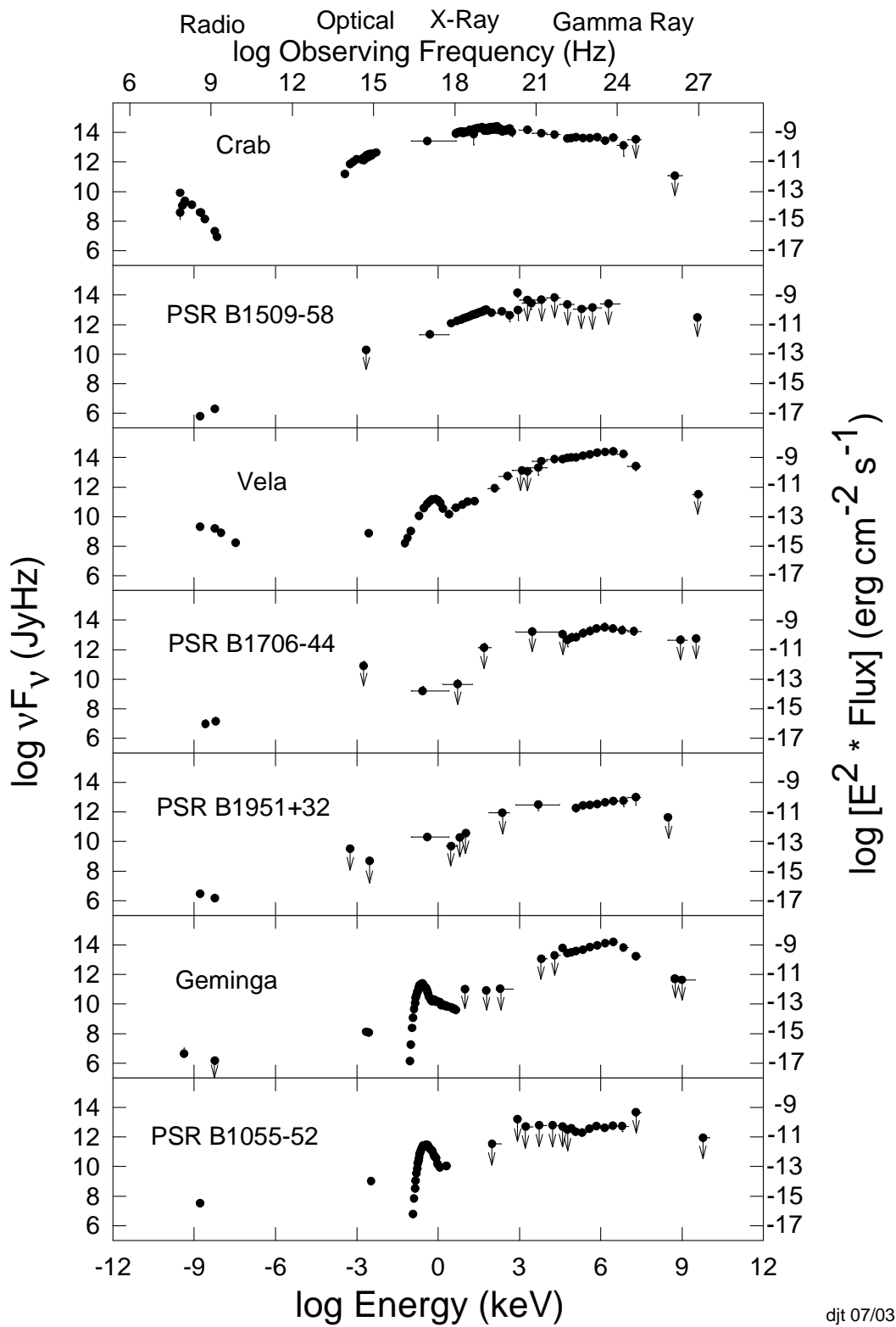}
\caption{Multiwavelength spectra of seven gamma-ray pulsars.  Updated from Thompson et al. (2004)}
\label{fig:figure9}
\end{figure}

Although the pulsations identify sources as rotating neutron stars, the observed energy spectra reflect the physical mechanisms that accelerate charged particles and help identify interaction processes that produce the pulsed radiation.  Broadband spectra for the seven highest-confidence gamma-ray pulsars are shown in Figure 9.  The presentation in $\nu$F$_{\nu}$ format (or E$^2$ times the photon number spectrum) indicates the observed power per frequency interval across the spectrum.  In all cases, the maximum power output is in the gamma-ray band. Other noteworthy features on this figure are:

\begin{itemize}
\item 
 The distinction between the radio emission (which originates from a coherent process) and the high-energy emission (probably from individual charged particles in an incoherent process) is visible for some of these pulsars, particularly Crab and Vela. 
 
\item Vela, Geminga, and B1055$-$52 all show evidence of a thermal component in X-rays, thought to be from the hot neutron star surface. 

\item The gamma-ray spectra of known pulsars are typically flat, with most having photon power-law indices of about 2 or less between 30 MeV and several GeV.  Energy breaks are seen in the 1-4 GeV band for several of these pulsars.  These changes in spectral index may be related to the calculated surface magnetic field of the pulsar.  The lowest-field pulsars have no visible break in the EGRET energy range; the existence of a spectral change is deduced from the absence of TeV detections of pulsed emission.  The highest-field pulsar among these, B1509$-$58, is seen only up to the COMPTEL energy band.

\item No pulsed emission is seen above 30 GeV, the upper limit of the EGRET observations, except for a recent detection of the Crab by the MAGIC telescope (Teshima, 2008).  The nature of the high-energy cutoff is an important feature of pulsar models. 

\item Although Figure 9 shows a single spectrum for each pulsar, the pulsed energy spectrum varies with pulsar phase.  A study of the EGRET data by Fierro et al. (1998) of the phase-resolved emission of the three brightest gamma-ray pulsars (Vela, Geminga, Crab) showed no simple pattern of variation of the spectrum with phase that applied to all three pulsars. A broadband study of the Crab by Kuiper et al. (2001) indicated the presence of multiple emission components, including one that peaks in the 0.1 - 1 MeV range for the bridge emission between the two peaks in the light curve. Improved measurements and modeling of the phase-resolved spectra of pulsars can be expected to be a powerful tool for study of the emission processes. 

\end{itemize}

The measured spectra can be integrated to determine the energy flux of each pulsar. Except for the Crab and PSR B1509$-$58, whose luminosity peaks lie in the $\sim$100 keV - 1 MeV range, the energy flux for the other gamma-ray pulsars is dominated by the emission above 10 MeV. The energy flux can be converted to an estimated luminosity by using the measured distance to the pulsar and an assumed emission solid angle.  For simplicity, I assume an emission into one steradian.  This value is unlikely to be the same for all pulsars but provides a simple reference point for comparison.  A significant uncertainty in such calculations is introduced by the distance estimate.    Table 2 summarizes these results for the ten pulsars.  P is the pulsar spin period in seconds.  $\tau$ is the estimated age of the pulsar.  $\dot E$ is the rate the pulsar is losing energy as it slows down.  F$_E$ is the energy flux seen at high energies (X-rays and gamma rays).   The distance d is given in kiloparsecs. L$_{HE}$ is the calculated high-energy luminosity of the pulsar.  $\eta$ is the efficiency for conversion of spin-down energy loss into high energy radiation. 

\begin{table}
\centering
\caption{\label{2} Summary Properties of the Highest-Confidence and Candidate Gamma-Ray Pulsars}
\bigskip
\begin{tabular}{lrrccccc}
\hline\hline
 Name  & P & $\tau$ & $\dot E$ & F$_E$ & d &  L$_{HE}$  & $\eta$ \\
     & (s) & (Ky) & (erg/s) &  (erg/cm$^{2}$s) & (kpc) & (erg/s)
& (E$>$1 eV)\\
\hline
Crab & 0.033 & 1.3 & 4.5 $\times$ 10$^{38}$ & 1.3 $\times$ 10$^{-8}$ & 2.0 & 5.0 $\times$ 10$^{35}$ &  0.001 \\
B1509$-$58 & 0.150 & 1.5 & 1.8 $\times$ 10$^{37}$ & 8.8 $\times$ 10$^{-10}$ & 4.4 & 1.6 $\times$ 10$^{35}$ &  0.009 \\
Vela & 0.089 & 11 & 7.0 $\times$ 10$^{36}$ & 9.9 $\times$ 10$^{-9}$ & 0.3 & 8.6 $\times$ 10$^{33}$ &  0.001 \\
B1706$-$44& 0.102 & 17 & 3.4 $\times$ 10$^{36}$ & 1.3 $\times$ 10$^{-9}$ & 2.3 & 6.6 $\times$ 10$^{34}$ &  0.019 \\
B1951+32 & 0.040 & 110 & 3.7 $\times$ 10$^{36}$ & 4.3 $\times$ 10$^{-10}$ & 2.5 & 2.5 $\times$ 10$^{34}$ &  0.007 \\
Geminga  & 0.237 & 340 & 3.3 $\times$ 10$^{34}$ & 3.9 $\times$ 10$^{-9}$ & 0.16 & 9.6 $\times$ 10$^{32}$ &  0.029 \\
B1055$-$52 & 0.197 & 530 & 3.0 $\times$ 10$^{34}$ & 2.9 $\times$ 10$^{-10}$ & 0.72 & 1.4 $\times$ 10$^{33}$ &  0.048 \\
\hline
B1046$-$58 & 0.124 & 20 & 2.0 $\times$ 10$^{36}$ & 3.7 $\times$ 10$^{-10}$ & 2.7 & 2.6 $\times$ 10$^{34}$ &  0.013 \\
B0656+14  & 0.385 & 100 & 4.0 $\times$ 10$^{34}$ & 1.6 $\times$ 10$^{-10}$ & 0.3 & 1.3 $\times$ 10$^{32}$ &  0.003 \\
J0218+4232 & 0.002 & 460,000 & 2.5 $\times$ 10$^{35}$ & 9.1 $\times$ 10$^{-11}$ & 2.7 & 6.4 $\times$ 10$^{33}$ &  0.026 \\
\hline
\end{tabular}
\end{table}

One trend, first noted by Arons (1996), can be derived from this table.  Figure 10 shows the efficiency of each pulsar as a function of the open field line voltage, the potential that can be developed by the rotating neutron star.  The efficiency increases as the open field line voltage decreases.  The implication of this figure is that there must be a limit to gamma-ray production by pulsars, because the efficiency shown in Figure 10 is approaching 1.  Pulsars with lower open field line voltage must either turn off or reach some saturation value.

 \begin{figure}
\centering
\includegraphics [height = 3.5 in.] {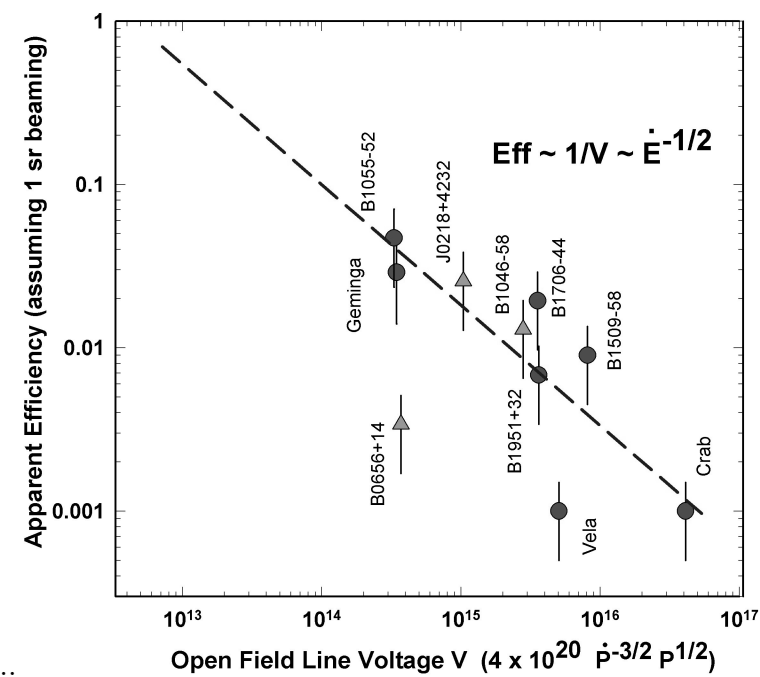}
\caption{Calculated pulsed high-energy (X-ray and gamma ray) efficiencies of the known and candidate gamma-ray pulsars, as a function of the open field line voltage. Circles: high-confidence gamma-ray pulsars. Triangles: lower-confidence gamma-ray pulsars. }
\label{fig:figure10}
\end{figure}

Despite trends such as the one shown in Figure 10, each one of the nine high-confidence and lower-confidence pulsars seen by EGRET has at least one unique feature.  Some examples:

\begin{itemize}
\item The Crab is the only gamma-ray pulsar to have its light curve aligned with the light curves seen at all other wavelengths.  This is the youngest gamma-ray pulsar and the one with the largest spin-down luminosity.  It is also the least efficient of these in converting spin-down luminosity into high-energy radiation. 
\item PSR B1951+32 is the only one of these to have a spectrum extending to the limits of the EGRET energy range with no evidence of a spectral cutoff.  Of the high-confidence pulsars, it is the only one not bright enough to appear as a source in the 3EG catalogue.
\item PSR B1706$-$44 is unique in having a gamma-ray spectrum with two power laws and a change of slope at about 1 GeV (compared to spectral cutoffs seen in others). 
\item Geminga is the only radio-quiet gamma-ray pulsar found by EGRET. 
\item PSR B1055$-$52 shows the highest efficiency of any of these pulsars for conversion of spin-down energy into gamma radiation.
\item PSR B1046$-$58 is the only one of these with no pulsed X-ray counterpart.
\item PSR J0218+4232 is the only millisecond pulsar with apparent gamma-ray emission.  It is also the only one of these pulsars with another candidate gamma-ray source (a blazar) close enough spatially to cause source confusion. 
\end{itemize}

Despite the wide range of information about these pulsars from their timing properties and their measured gamma-ray characteristics, the fact that a special feature can be found for each one limits our ability to draw broad conclusions. 

Stimulated by these observations of gamma-ray pulsars, theorists have carried out extensive modeling of these high-energy neutron stars, but without reaching a consensus on where the particles are being accelerated in the star's magnetosphere or how these particles interact to produce the gamma radiation.  Some examples of theoretical high-energy pulsar modeling are Sturrock (1971), Ruderman and Sutherland (1975), Romani (1996), Harding and Muslimov (2005), Cheng and Zhang (1998), and Hirotani (2008).  A useful overview of such theoretical work and the implications for pulsar population studies is given by  Harding \etal (2007). 

In addition to the gamma-ray pulsars identified by their periodic emission, there are several potential associations of known pulsars with EGRET sources. Some radio pulsars discovered after the end of the CGRO mission are positionally consistent with 3EG sources and have enough energy to power the observed gamma radiation (e.g. Kramer et al. 2003, Halpern et al. 2001).  A related case is 3EG J1835+5918, with properties suggesting an association with the isolated neutron star RX J1836.2+5925 (Mirabal and Halpern 2001; Reimer \etal 2001; Halpern \etal 2002).  Because gamma-ray pulsars appear to have significant timing noise, searching for pulsations in the EGRET data by extrapolating timing solutions back in time involves too many trials to produce high-confidence results, as illustrated by the unsuccessful search for PSR J2229+6114, found in the error box of 3EG J2227+6122 (Thompson \etal 2002)

\subsection{Binary sources}

The EGRET data showed some indication of gamma radiation from binary sources, although the case is far from certain.  Cen X-3 , an accretion-powered X-ray binary, may have shown a flare during an EGRET observation, including some evidence of variability at the 4.8 second spin period of the neutron star (Vestrand \etal  1997).   Two sources in the 3EG catalogue,  3EG J0241+6103    and 3EG J1824$-$1514 are positionally consistent with high-mass X-ray binary systems (HMXB) often placed in the microquasar class, LSI +61$^\circ$303 (Kniffen \etal 1997; Tavani \etal 1998) and LS 5039 (Paredes \etal 2000) respectively.  The detection of TeV radiation from both these HMXBs, showing orbitally modulated emission  for LSI +61$^\circ$303 (Albert \etal 2006) and periodic emission for LS 5039 (Aharonian \etal 2007), indicates that such sources can accelerate particles to energies well beyond those needed to produce gamma rays in the EGRET energy band.  The evidence that the EGRET sources are actually these binary systems remains largely circumstantial, however, and searches for other binaries in the EGRET data have been unsuccessful. 

 \subsection{Other Galactic sources}
 
 Most of the 3EG sources along the Galactic plane remain unidentified, including one that flared up to be the second brightest source in the gamma-ray sky for less than one month (Tavani \etal 1997).  Two general approaches have been followed in order to shed light on the possible nature of these sources:
 
 \begin{itemize}
\item The characteristics of the Galactic EGRET sources as a class, based on their spatial and spectral properties, was investigated by several authors.  Starting with the earlier 2EG catalogue (Thompson \etal 1995), Mukherjee \etal (1995), Kanbach \etal (1996) and Merck \etal (1996) compared the characteristics of unidentified sources with Galactic tracers.  McLaughlin \etal (1996) developed a method for characterizing gamma-ray source variability and found that some of the catalogued Galactic sources appear to be variable. \"Ozel and Thompson (1996) constructed log N-log S distributions for 2EG sources, showing that the number of unidentified sources N with flux greater than S at high Galactic latitudes had an isotropic distribution, consistent with being either quite local or extragalactic. Similar analyses were carried out for the 3EG catalogue. Reimer and Thompson (2000) and Bhattacharya \etal (2003) studied the log N-log S distributions. Spatial-statistical considerations and variability studies suggest there is a population of Galactic and variable GeV gamma-ray emitters among the unidentified EGRET sources (Nolan \etal 2003).  A population of steady gamma-ray sources with different properties from those close to the Galactic Plane,  possibly associated with the nearby Gould Belt complex of gas and stars, was suggested by Gehrels \etal (2000).  As noted by Cassandjian and Grenier (2008), many of these may be gas clouds that were not modeled in the EGRET analysis. A valuable summary of 3EG source characteristics, along with some cautions about limitations, is given by Reimer (2001).  The strongest conclusion from these many studies was that the EGRET Galactic sources comprise more than one population. 

\item Correlations of known Galactic populations with the positions of the EGRET sources offered another approach to trying to discern their nature. Using various statistical techniques, various authors found indications of associations of EGRET sources with
star forming regions or groups of hot and massive stars (e.g. Kaaret and Cottam 1996; Romero \etal 1999), supernova remnants (e.g. Sturner and Dermer 1995; Esposito \etal 1996), or pulsar wind nebulae (e.g. Roberts, Romani and Kawai 2001).  Torres \etal (2003) provide an excellent summary of the observational and theoretical possibilities for supernova remnants as EGRET sources.  
  Pulsar populations may also explain a fraction of the Galactic unidentified sources  (Yadigaroglu and Romani 1997).  All these classes of sources are plausible, because they have the potential to accelerate particles to high energies in an environment in which the particles can interact to produce gamma rays.  There are several issues with these analyses:
   \begin{itemize}
  \item  All these sources tend to be located in the same regions.  Particularly considering the possibility that more than one type of gamma-ray source may be present, separating the classes is difficult.
  \item There is no unique spectral or timing signature for most of these sources, in the absence of pulsations or orbital periods (which have not been found). 
  \item The EGRET error boxes are too large to make unique associations possible.  In fact, none of these approaches found a clear example of one source that would serve as a prototype for the class.  
\end{itemize}
Ultimately, all these efforts proved valuable in pointing to young, active Galactic objects as likely gamma-ray sources. 
\end{itemize}

\section{Extragalactic gamma-ray sources}

As seen in figure 7, EGRET sources are seen in all directions in the sky.  The detection of the nearby quasar 3C273 by COS-B (Swanenburg \etal 1978) had long suggested that extragalactic sources would be a significant component of the high-energy sky (e.g. Bignami \etal 1979).  

 \subsection{Blazars}
 
 Just three months after the launch of the Compton Observatory, a Target of Opportunity pointing (requested by the OSSE team in search of emission from a supernova in the Virgo cluster) produced the first big surprise in the EGRET data.  The observation that was expected to produce an observation of 3C273 was instead dominated by a very bright source about 10 degrees away, positionally consistent with a more distant quasar, 3C279, with a redshift z = 0.536 (Hartman \etal 1992).  This detection had two immediate implications:
 
  \begin{itemize}
  \item  The gamma-ray sky is variable, at least on long time scales, because COS-B had not seen this source.
  \item The idea of nearby AGN as candidate gamma-ray sources was at best incomplete.  The fact that 3C279 was part of the blazar subclass of AGN, thought to be powered by supermassive black holes and having powerful jets of particles and radiation pointed toward the Solar System (Blandford and Rees 1978; Blandford and K\"onigl 1979), suggested that gamma rays might be a valuable probe of jet sources. 
\end{itemize}

The recognition of short-term variability of 3C279 (on a scale of days, Kniffen \etal 1993) reinforced the idea that jets were a likely source of the EGRET-detected gamma rays. The rapid variability required a compact emitting region, and such a region should be opaque to gamma rays due to gamma-gamma pair production, the interaction of a high-energy gamma ray with a lower-energy photon to produce an electron-positron pair, $\gamma$ + $\gamma$ = e$^+$ + e$^-$, unless most of the photons are moving in the same direction, as in a jet.  Some example calculations of this process are those of Mattox \etal (1993) and Sikora \etal (1994).   The ultimate confirmation of such gamma-ray sources as blazars came with multiwavelength campaigns that found correlated variability of gamma-ray flares with flares seen at other wavelengths.  The first example was a flare seen in PKS 1406$-$076 in January 1993, shown in Figure 11, where an optical flare was seen nearly  simultaneous with the gamma-ray flare (Wagner \etal  1995). Such correlated variability is a valuable source of information about how such jets are formed, how they are collimated, and how they carry energy. 
 
\begin{figure}
\centering
\includegraphics [height = 3.5 in.] {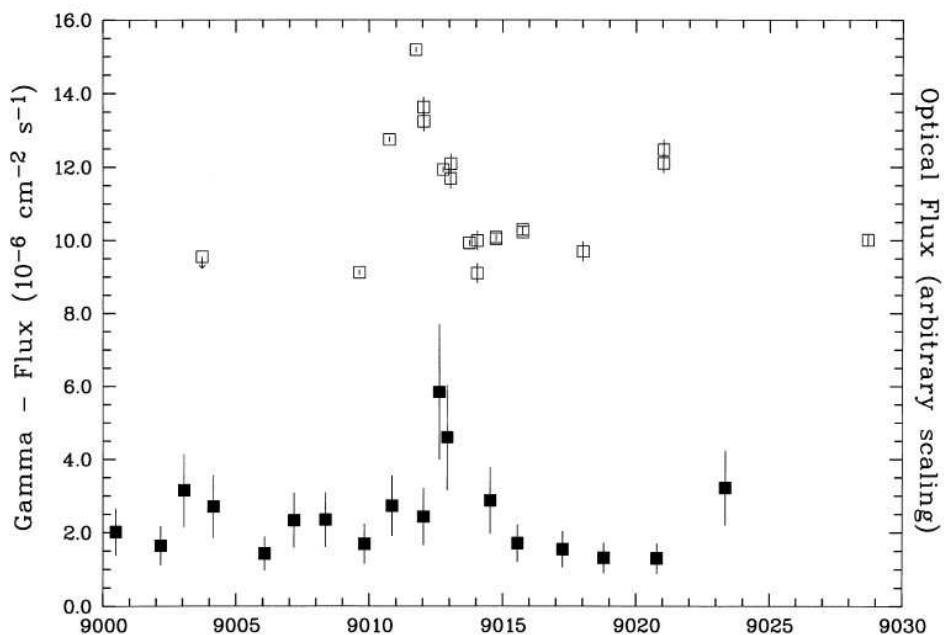}
\caption{Optical and gamma-ray flare in  PKS 1406$-$076 (Wagner \etal  1995) }
\label{fig:figure11}
\end{figure}
 
 \begin{figure}
\centering
\includegraphics [height = 4.5 in.] {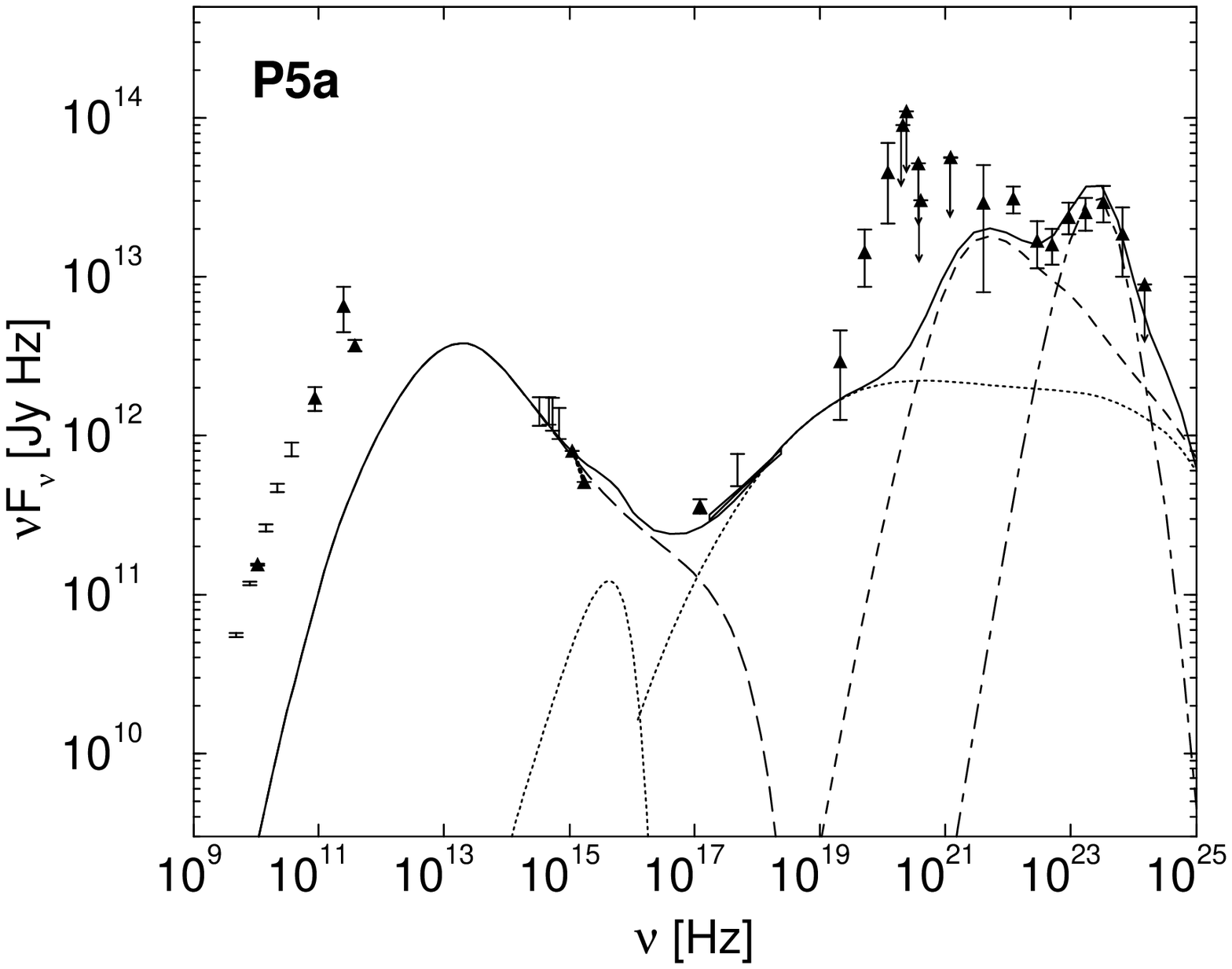}
\caption{Spectral Energy Distribution for 3C279 during a bright state in January 1996 (just before a large flare) with models (Hartman \etal  2001a). Modeled components, from left to right in the figure: synchrotron radiation, thermal radiation from the accretion disk, synchrotron-self-Compton radiation, Compton radiation from scattering of accretion disk photons, and Compton radiation from scattering of photons from gas clouds. }
\label{fig:figure12}
\end{figure}

 Continuing EGRET observations revealed a series of bright, well-localized sources positionally consistent with prominent blazars.  Many of these were known as Optically Violently Variable (OVV) quasars or BL Lacertae (BL Lac) objects.  Although the term ``blazar'' is not uniquely defined, it typically encompasses those Active Galactic Nuclei with the following characteristics: radio-loud, with flat radio spectrum; significant polarization in optical and/or radio; significant variability.  Blazars are seen across the electromagnetic spectrum, with a characteristic two-peak Spectral Energy Distribution (SED), as illustrated in Figure 12 (Hartman \etal 2001a).  At lower frequencies, from radio to optical or sometimes X-rays, the emission is thought to be dominated by synchrotron radiation of high-energy electrons in the jet.  The upper peak, extending from X-rays upward, is thought to be primarily inverse Compton scattering of low-energy photons by the same population of high-energy electrons that produces the lower-energy synchrotron radiation.  The source of the photons to be upscattered can be the synchrotron radiation itself (Synchrotron Self-Compton) or some outside source of photons (External Compton).  Extensive modeling of blazars has evolved following the EGRET discoveries.  An example of such modeling is shown in Figure 12.   In many cases, the gamma-ray emission is the dominant observable output of these blazars.

  \begin{figure}
\centering
\includegraphics [height = 8.0 in.] {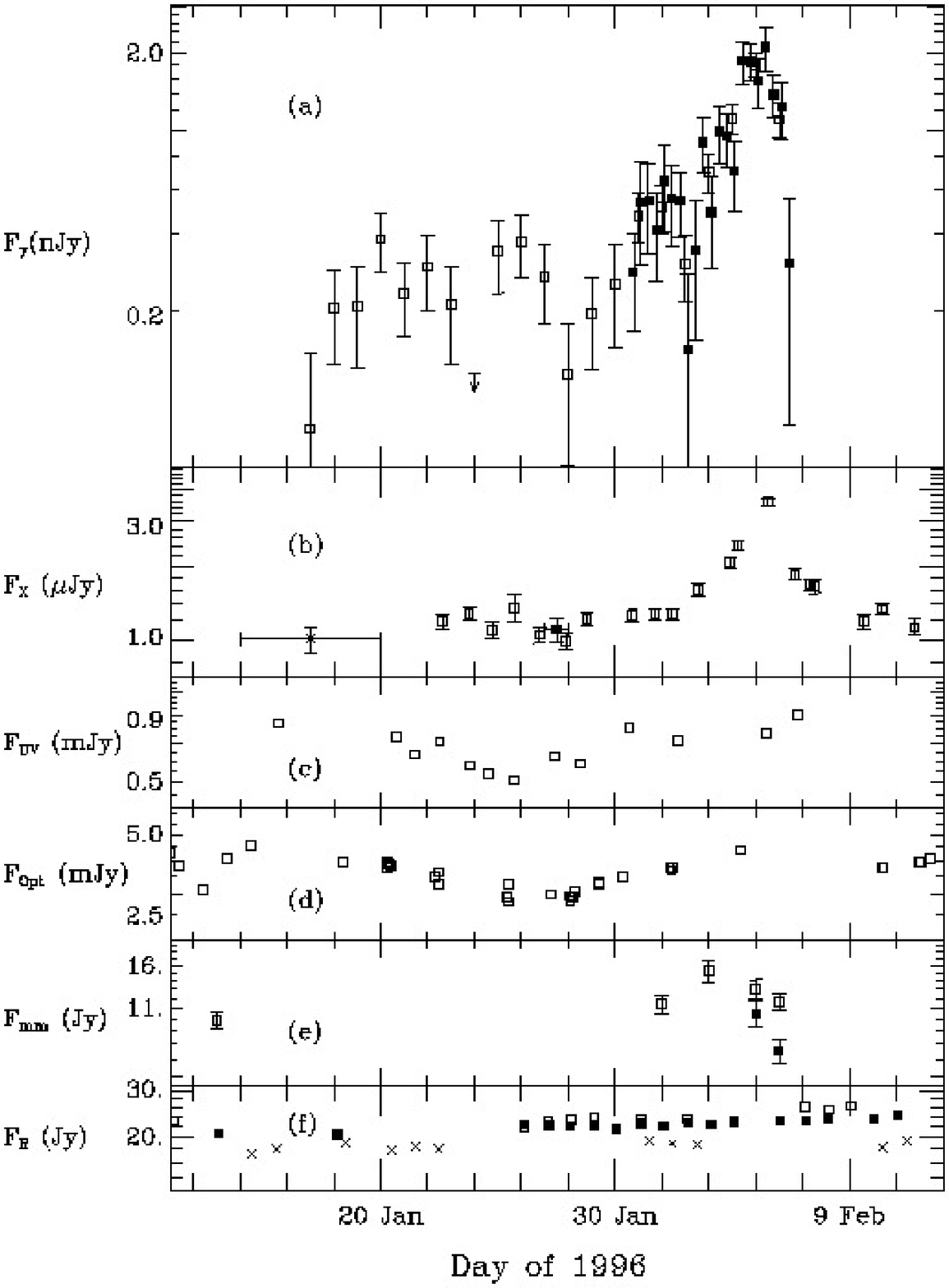}
\caption{Correlated multiwavelength emission from 3C279 during the large flare of January-February 1996 (Wehrle \etal 1998) }
\label{fig:figure13}
\end{figure}

 Some of the EGRET blazars of particular interest are:
 \begin{itemize}
  \item  3C279.  This first blazar recognized by EGRET was prominent many times during the CGRO mission.  It was the first to be involved in a concentrated multiwavelength campaign (Maraschi \etal 1994).  A later multiwavelength campaign in January-February 1996, shown in Figure 13, captured a dramatic flare, seen at multiple wavelengths (Wehrle \etal 1998).  The gamma-ray flux increased by over a factor of 10 during this flare. As a point of interest, the limited optical coverage of this flare came about because most of the optical astronomers who monitor this source were attending a blazar workshop at the time and were not at their telescopes. 
  \item PKS 1622$-$297.  Located not far from the Galactic Center region, this lesser-known blazar was not seen in any of the first 17 EGRET observations of this region, then flared up in the Summer of 1995 to be the brightest gamma-ray source seen by EGRET to that time, and the one showing the fastest time variability, doubling in flux in less than 8 hours, with the limitation being just the photon statistics (Mattox \etal 1997)
\item  Mkn 421. This well-known BL Lac object appeared in an early EGRET observation (Lin \etal 1992).  Its detection helped stimulate the TeV gamma-ray astronomy community, and it became the first blazar detected at TeV energies (Punch \etal 1992). 
\item PKS 0528+134.  This blazar, actually the first one that appeared in the EGRET data although not identified until later, is one of at least five seen by EGRET with redshift z $>$ 2.0 (Hunter \etal 1993).  Although it underwent a strong flare during one 1993 observation (Figure 14, Mukherjee \etal 1999),  it was also seen to be quite stable during other observations, even in short-timescale investigations (Wallace \etal  2000). 
\item PKS 2255$-$282.  After the completion of the third EGRET catalogue, this blazar was the only new one to appear with high significance in the EGRET data  (Macomb \etal 1999).  The bright gamma-ray flare appeared in early 1998 following a period of rising flux in the submillimeter band (Tornikoski \etal 1999). 
\end{itemize}

  \begin{figure}
\centering
\includegraphics [height = 4.0 in.] {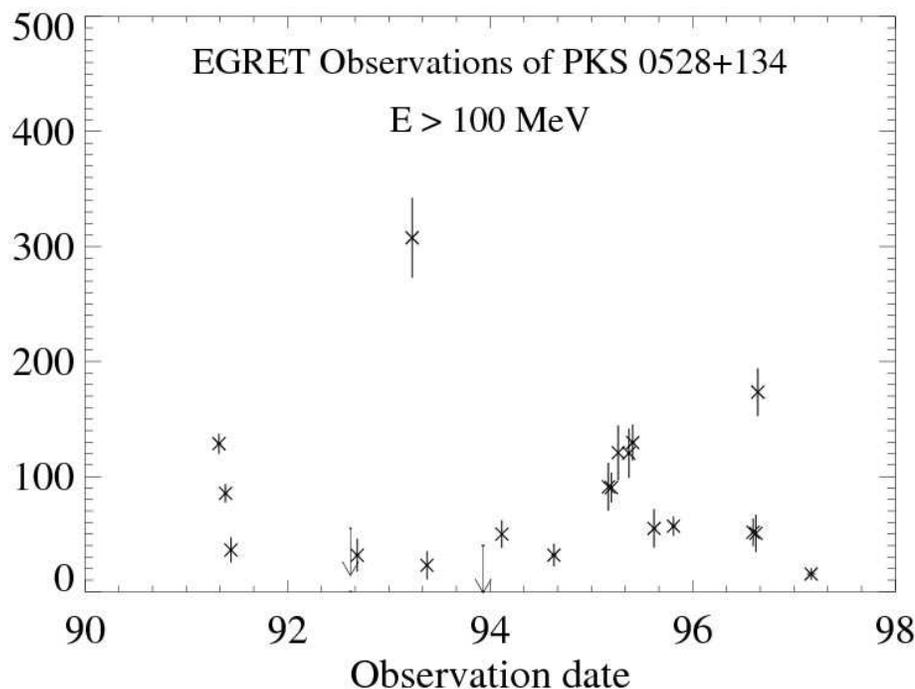}
\caption{Light curve for PKS 0528+134 during the nine-year life of EGRET (Mukherjee \etal 1999) }
\label{fig:figure14}
\end{figure}

The EGRET discoveries of numerous highly-variable gamma-ray blazars provided a stimulus to this field of study and to other studies of Active Galactic Nuclei.  Terminology in this field has also evolved, and blazars are now usually classified as Flat-Spectrum Radio Quasars (FSRQs), Low-frequency Peaked BLLac Objects (LBLs), and High-frequency Peaked BLLac Objects (HBLs).  With blazars as the most numerous class of identified gamma-ray sources, they offered opportunities for multiwavelength comparisons and classifications. Some examples:

\begin{figure}
\centering
\includegraphics [height = 4.5 in.] {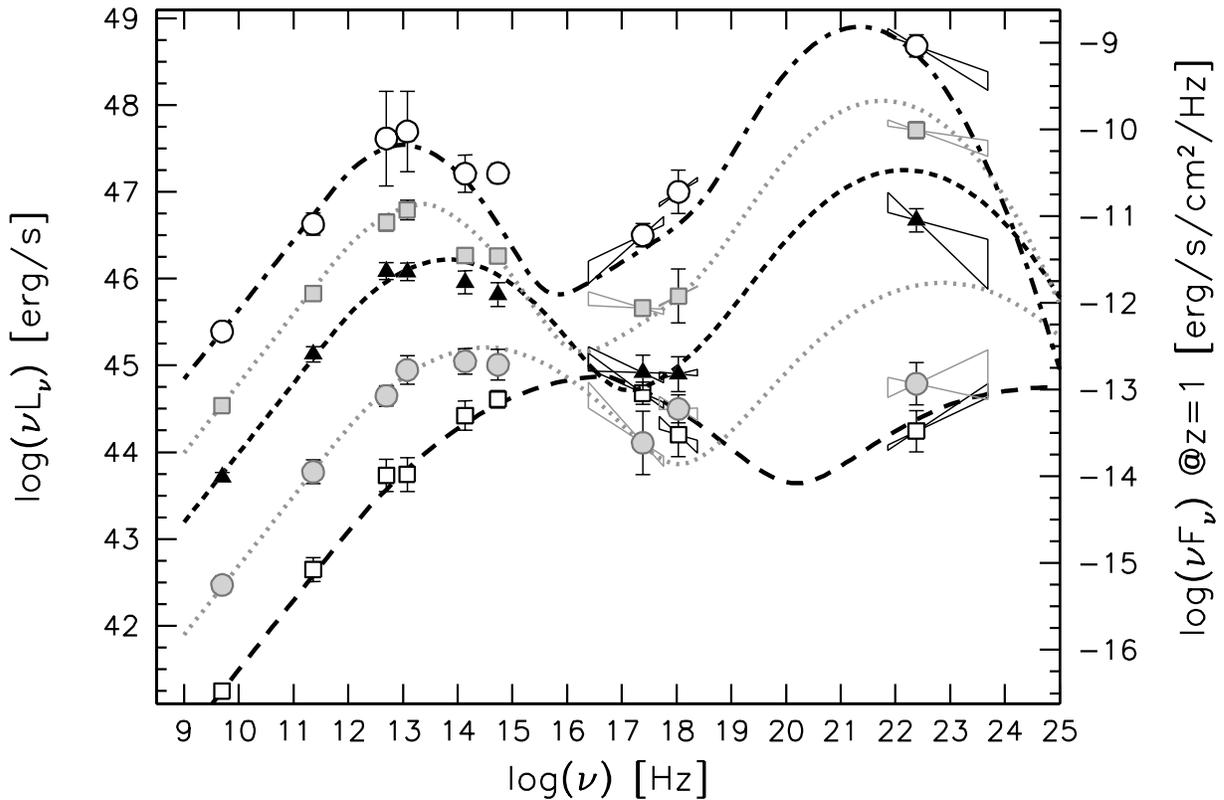}
\caption{The blazar sequence concept of Fossati \etal (1998).  These Spectral Energy Distributions for blazars show a multiwavelength pattern  in which FSRQs (toward the top) have higher luminosity and peaks at lower energy, while LBLs and HBLs have lower luminosity but peaks at higher energy.  The EGRET gamma-ray spectra are to the right in this figure. Figure courtesy of G. Fossati. }
\label{fig:figure15}
\end{figure}

 \begin{itemize}
 \item A number of prominent blazars were not seen by EGRET despite significant exposures (von Montigny \etal  1995a).
  \item  An effort was made to find a unified model for AGN based on geometry: the orientation of the black-hole/accretion-disk/torus/jet system relative to the viewing direction (Urry and Padovani, 1995).  The beamed nature of the EGRET gamma radiation, associated with apparent superluminal motion seen in the radio, is an important aspect of this scheme, because it emphasizes that blazars must have jets aligned close to our line of sight.  
  \item A comparison of spectral energy distributions led Fosatti (1998) to suggest the ``blazar sequence'' in which higher-luminosity blazars (usually FSRQs) have their synchrotron and Compton peaks at lower energies while lower-luminosity blazars (LBLs and HBLs) have peaks at increasingly higher energies as their luminosity decreases (see Figure 15).
\item  Very Long Baseline Interferometry (VLBI) radio studies have been used to construct time-resolved images of the jets associated with the EGRET blazars.  In addition to finding that many of these blazars have apparent superluminal motions (a relativistic illusion produced by the fact that the jet is directed close to our line of sight), a study of a sample of bright blazars suggests that those blazars seen in gamma rays tend to have higher jet Lorentz factors than those blazars not seen by EGRET (Figure 16, Kellerman \etal 2004).
\item Another important result from VLBI studies is an indication that gamma-ray flares seen with EGRET occur at approximately the same time that new components of radio emission (``knots'') emerge from the core of the AGN (Jorstad \etal 2001). This result suggests that the gamma radiation is produced in the parsec-scale region of the jet rather than closer to the black hole at the center of the AGN. 
\end{itemize}

Summaries of the observational properties of the EGRET blazars were given by von Montigny \etal (1995b) and Mukherjee \etal (1997).   Mukherjee (2001) reviewed the EGRET blazar results after the end of the CGRO mission.  Efforts to search the Third EGRET catalogue systematically for blazar-like counterparts were carried out by Mattox \etal (2001) and by Sowards-Emmerd \etal  (2004).  A number of plausible associations were added to the ones shown in the catalogue.  Figure 17 shows a map of such associations.  
 
 Although many studies of these blazars provided useful insights into jets and other AGN features, others found puzzling results.  The Sowards-Emmerd \etal  results concluded that some of the high-Galactic-latitude EGRET sources did not have clear identifications with blazar-like sources.  Nandikotkur \etal (2007) carried out a detailed study of gamma-ray blazar variability using the entire EGRET data set and found a variety of relationships between flux and spectral hardening, without a simple pattern.  Hartman \etal (2001b) concluded that at least one episode of variability of 3C279 was not correlated with optical or X-ray variation.  No consensus has emerged about whether the jets are primarily dominated by electrons or by protons.
 
 \begin{figure}
\centering
\includegraphics [height = 3.5 in.] {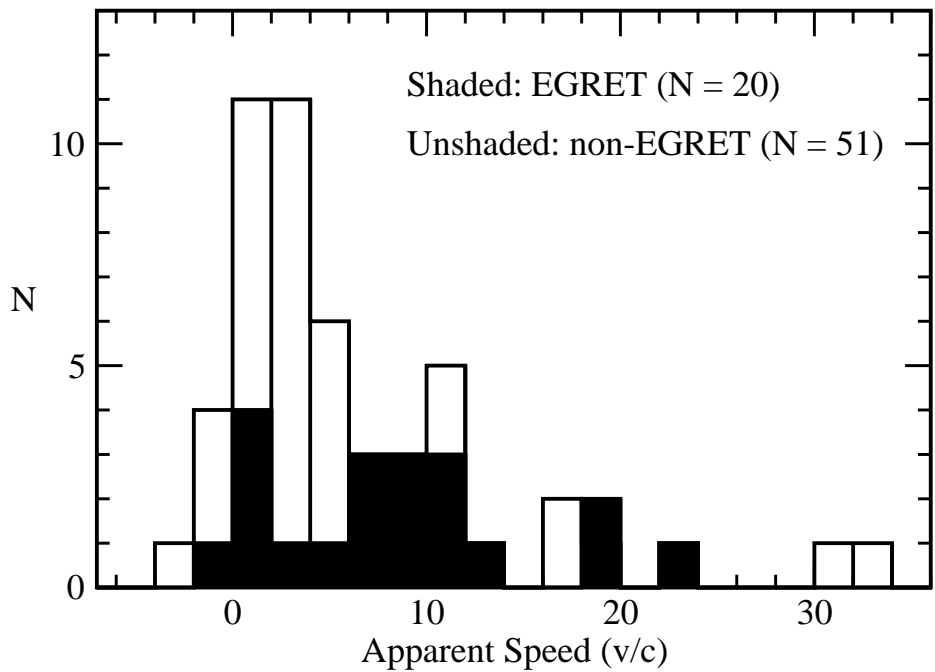}
\caption{Histogram of speeds of the fastest component of VLBI-imaged blazar jets (Kellerman \etal 2004).  The EGRET blazars have a higher average speed than the others.  }
\label{fig:figure16}
\end{figure}

\begin{figure}
\centering
\includegraphics [height = 3.5 in.] {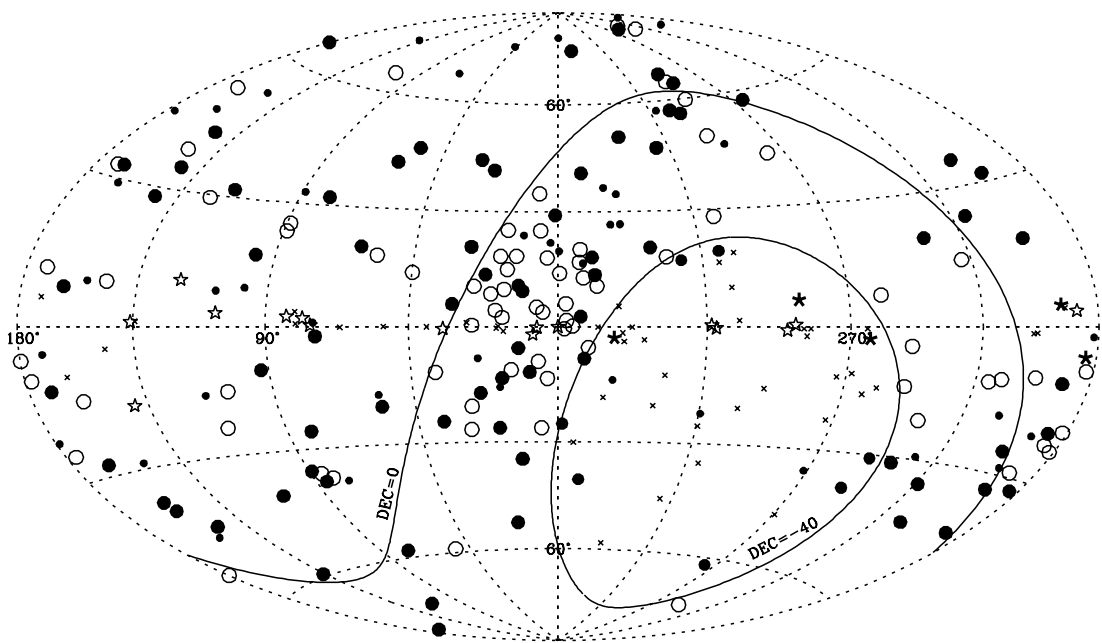}
\caption{New candidate identifications for EGRET sources (Sowards-Emmerd \etal 2004).  The sources are the same as the 3EG catalogue (Figure 7).  The filled circles show potential blazar associations.  The sources shown by open circles do not have blazar counterparts.  This analysis did not extend to declinations south of -40$^\circ$, shown by a solid contour.  Crosses show sources that were not classified.  }
\label{fig:figure17}
\end{figure}

 \subsection{Other Galaxies}
 
 As noted in the previous section, not all EGRET sources at high Galactic latitudes are identified as blazars.  Other potential extragalactic sources are discussed below. 
 
\subsubsection{Local Galaxies}

The same type of cosmic ray interaction processes that  operate in our Galaxy are likely to operate in other normal galaxies, although most of these are too far away to be detectable.  EGRET was able to set only upper limits on gamma radiation from the Andromeda Galaxy, M31, for example (Sreekumar \etal 1994). One exception is the Large Magellanic Cloud (LMC), reported as an extended gamma-ray source by Sreekumar \etal (1992).  Based on radio and optical observations, the LMC is thought to have a cosmic ray population similar to that of the Milky Way.  The source seen by EGRET was consistent in flux and spatial extent with resulting from cosmic ray interactions in the LMC.  A map of the EGRET emission and a comparison to the radio emission that traces the gas content of the LMC is shown in Figure 18.

   \begin{figure}
\centering
\includegraphics [height = 3.5 in.] {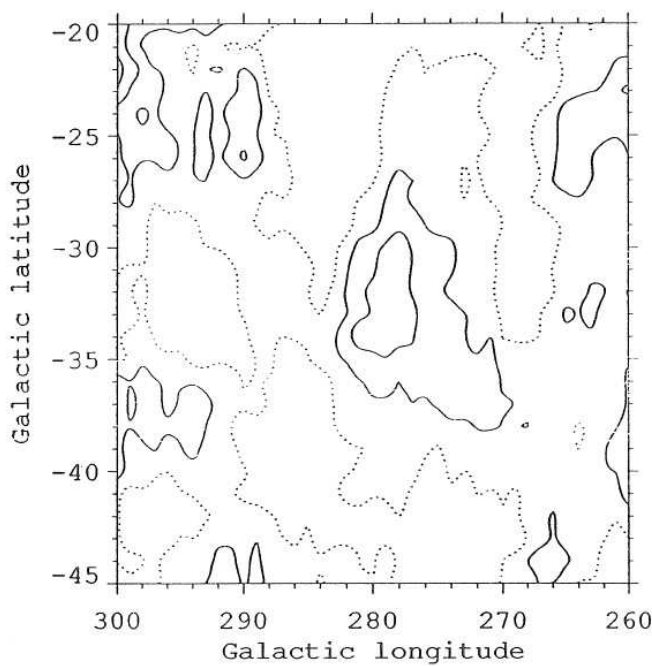}
\caption{EGRET gamma radiation from the Large Magellanic Cloud region (Sreekumar \etal 1992).  The contours of gamma-ray intensity show an extent consistent with that seen at radio frequencies. }
\label{fig:figure18}
\end{figure}

By contrast, Ginzburg (1972) predicted that the Small Magellanic Cloud (SMC) would not be detectable at the sensitivity of EGRET, because that galaxy is thought to be unable to sustain a local cosmic ray population.   The absence of significant gamma radiation from the SMC was a critical test of whether cosmic rays are confined to galaxies or whether they are more universal.  The SMC would only be a gamma-ray source if cosmic rays extended beyond stable galaxies.  The upper limit from EGRET (Sreekumar \etal  1993) provided the evidence to prove Ginzburg's hypothesis that cosmic rays are confined to galaxies. 

\subsubsection{Radio Galaxies and Others}

Evidence for gamma-ray emission from other extragalactic sources is less secure.  The nearest large radio galaxy, Cen A, may have been seen by EGRET.  There is a catalogued EGRET source positionally consistent with Cen A, and the energy spectrum appears to be a continuation of the spectrum seen at lower energies (Sreekumar \etal 1999).  In the absence of any variability correlated with other wavelengths, however, this identification is not certain.  A similar situation exists for two other radio galaxies. NGC 6251 is located in an EGRET source error box and is a plausible, but not definite, identification (Mukherjee \etal 2002), and 3C111 has been suggested as a possible counterpart to another EGRET source (Sguera \etal 2005).
  
 Searches for other extragalactic populations include:
 
  \begin{itemize}
  \item  Upper limits on some starburst galaxies were given by Sreekumar \etal (1994).
  \item Searches for EGRET gamma rays from clusters of galaxies were reported by Reimer \etal   (2003), who also provide a critical analysis of some suggestions of cluster emission in the EGRET data. 
  \item Stacking analyses gave upper limits in examining possible low-level contributions from radio galaxies (Cillis \etal 2004) and luminous infrared galaxies (Cillis \etal 2005). 
\end{itemize}

 \subsection{Diffuse Extragalactic Radiation}
 
 Finding a signal of diffuse extragalactic gamma radiation, sometimes called the extragalactic gamma-ray background, is probably the most difficult analysis challenge for a gamma-ray telescope, because this emission is what remains after all foreground constituents of the radiation are subtracted.  For the EGRET analysis, the foreground included three principal components, dealt with in separate steps by Sreekumar \etal (1998):
 
  \begin{itemize}
  \item  Individual sources, as identified for the EGRET catalogue work, were removed from the analysis;
  \item A residual contribution from the bright Earth limb was eliminated by removing all photons whose arrival directions were within 4 times the Point Spread Function  from the limb (for a given energy), enlarged from the standard EGRET cut of 2.5 times the PSF from the limb; 
  \item The Galactic diffuse component was taken into account, based on the standard model developed for EGRET analysis (see section  4). 
\end{itemize}
 
 The third of these steps was the most critical, because the Galactic diffuse emission exceeds any extragalactic diffuse radiation everywhere on the sky, including the Galactic poles.  The basic approach was to compare the observed radiation to that expected from the Galactic model and then to extrapolate to a condition of no Galactic emission.  Figure 19 shows such a fit for the standard EGRET energy bands.
 
  \begin{figure}
\centering
\includegraphics [height = 8.0 in.] {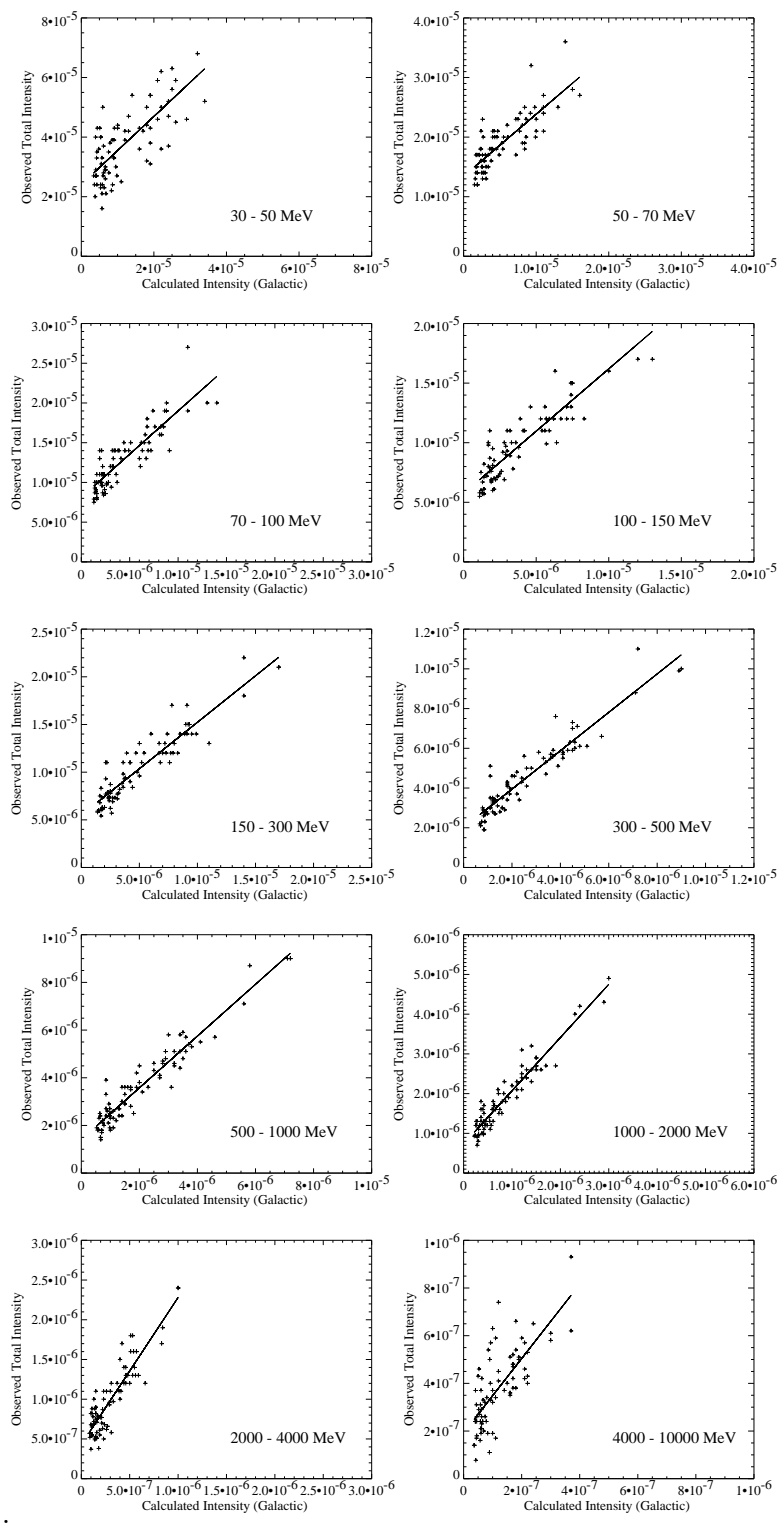}
\caption{Observed gamma ray flux in many high-latitude locations, comparing that observed (Y-axis) to the flux expected from the EGRET model of the Galactic diffuse emission (Sreekumar \etal 1998). The correlation is strong, and the offset from zero represents the diffuse emission.}
\label{fig:figure19}
\end{figure}

 This analysis produced a non-zero residual for all energies.  The resulting extragalactic diffuse spectrum was consistent with a power law with an index of $-$2.10+/-0.03.
 
The unresolved gamma radiation after removal of foreground contributions is certainly not all truly diffuse emission.  As with the X-ray background, some or all of this radiation results from individual sources that are too faint to be recognized by EGRET.  Primary candidates those from  the known extragalactic gamma-ray source classes:

 \begin{itemize}
  \item  Normal galaxies like the Milky Way must contribute.  Pavlidou and Fields (2002) estimate that up to 30\% of the emission at 1 GeV may result from such galaxies, with a lesser contribution at other energies.  With a sample of only the Milky Way and the LMC, however, any extrapolation must be considered highly uncertain. 
  \item Blazars are strong candidates to be the origin of much of the unresolved emission.  Many authors have estimated this contribution, and the results range from about 25\% to 100\% of the total (e.g. Stecker and Salamon 1996; Mukherjee and Chiang 1999; Chiang and Mukherjee 1998; M\"ucke and Pohl 2000). The principal challenges for this calculation are twofold: (1) Blazars seen by EGRET were almost always in a flaring state, and the duty cycle for such flaring remains highly uncertain; and (2) the evolution of blazar gamma-ray emission must be estimated.  The wide range of potential contributions to the diffuse radiation reflects the sensitivity of the calculation to the input assumptions. 
\end{itemize}

Beyond the contributions from known but unresolved sources, there is no shortage of other candidate mechanisms for producing a largely isotropic gamma-ray background on the scale observed by EGRET.  Dermer (2007) presents a recent summary. Some possibilities that have been discussed in recent years include: unresolved emission from  galaxy clusters (Ensslin \etal 1997), starburst galaxies (Thompson, Quataert and Waxman 2007), shock waves associated with large scale cosmological structure formation (Loeb and Waxman 2000; Miniati 2002), distant gamma-ray burst events (Casanova, Dingus and Zhang 2007),  annihilation of weakly interacting massive particles (WIMPs) (e. g. Ullio \etal 2002), and cosmic strings (Berezinsky, Hnatyk and Vilenkin 2001).
  
Strong \etal (2000) and Moskalenko and Strong (2000) have argued for a larger component of the Galactic diffuse emission from inverse Compton scattering of the Galactic plane photons and the cosmic microwave background.  Because the calculation of the extragalactic emission depends on correct subtraction of the Galactic contribution, this alternate approach would change the result on any determination of the extragalactic diffuse emission.   A new model of the Galactic diffuse emission (Strong \etal 2004a) has provided a new estimate of the extragalactic diffuse gamma radiation that is lower in flux and steeper than found by Sreekumar \etal (1998) The revised spectrum is not consistent with a power-law and shows some positive curvature.  Both results are shown in Figure 20.

 \begin{figure}
\centering
\includegraphics [height = 4.0 in.] {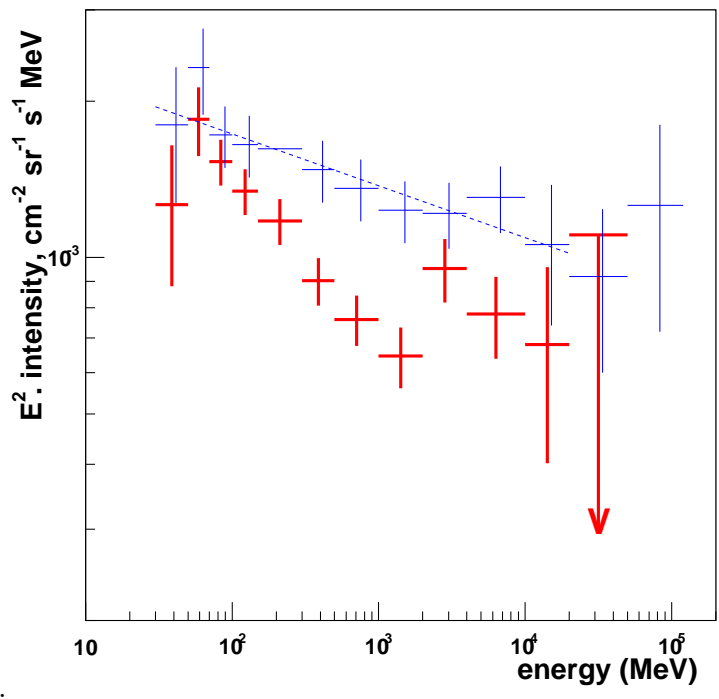}
\caption{Spectrum of the diffuse extragalactic gamma radiation.  Upper (lighter) data points, connected by dashed line: Sreekumar \etal (1998) analysis. Lower data points: Strong \etal (2004a) analysis.}
\label{fig:figure20}
\end{figure} 
   
  \subsection{Gamma ray bursts}
  
  Gamma-ray bursts (GRBs) have been described as the brightest explosions in the Universe,  radiating more energy in a few seconds than our sun will emit during its entire 10 billion year lifetime.  Much of this energy is seen as low-energy gamma rays, with typical energies in a range around 100 keV.  GRBs are thought to be caused by unusually powerful supernovae (collapsars) or possibly an effect of the merger of two neutron stars or a neutron star and a black hole (for a recent review, see Meszaros 2006).  During the Compton Observatory mission, BATSE recorded more then 2700 GRBs, averaging nearly one per day.  A small fraction of these bursts were also seen by EGRET, revealing important characteristics of GRBs. 

\begin{figure}
\centering
\includegraphics [width = 4.5 in., angle =270] {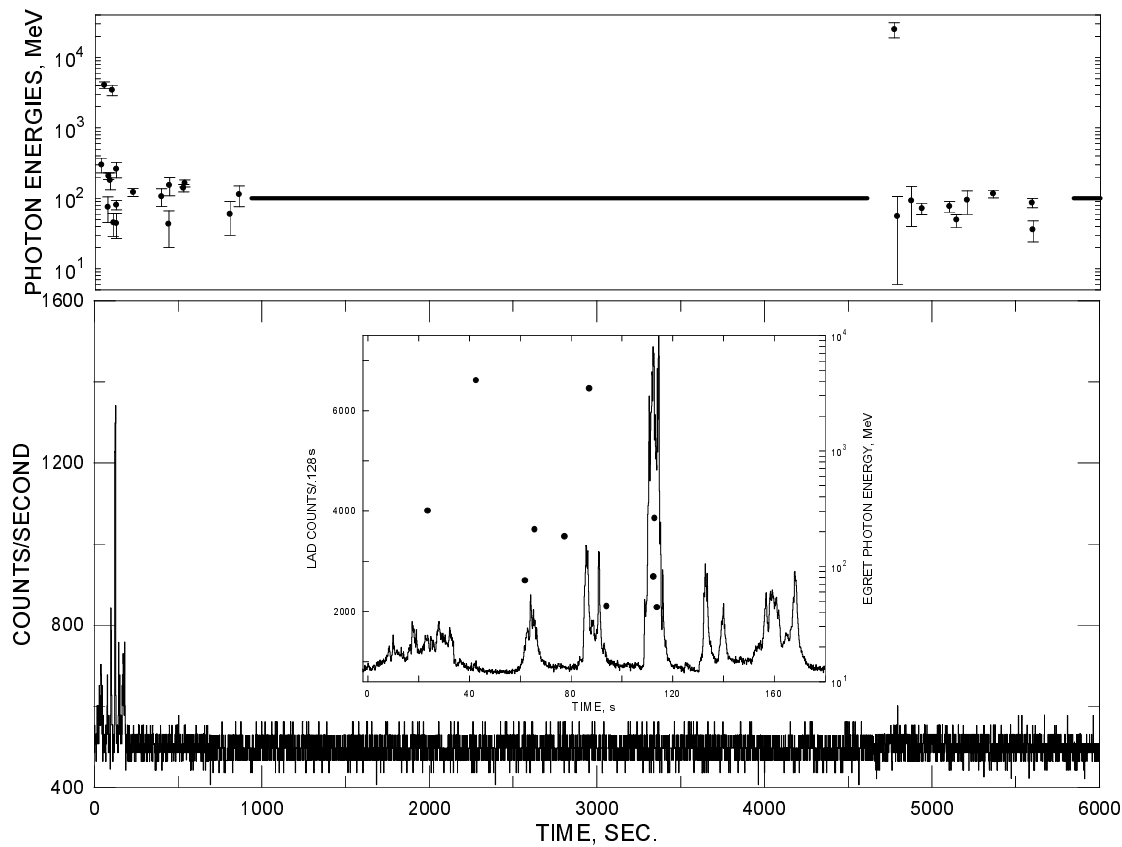}
\caption{The long-duration high-energy emission from the gamma-ray burst of 17 Feb. 1994 (Hurley \etal 1994). Upper panel: energies and arrival times of individual EGRET photons from the direction of the burst. The horizontal lines show periods with no data. Lower panel: Ulysses 25-150 keV counts rates during the same time period.  Inset: expanded plot of the early time of the burst, showing the EGRET photons compared to the BATSE light curve.  }
\label{fig:figure21}
\end{figure}

EGRET photons above 30 MeV were seen coincident with BATSE low-energy gamma-ray emission for at least 4 bursts (for a summary, see Dingus 2003).  Because these were the brightest BATSE bursts in the EGRET field of view and EGRET upper limits from other bursts do not constrain the results, it is possible that all GRBs have a high-energy component.  The combined energy spectrum shows no break up to 10 GeV, suggesting that the spectrum may extend to even higher energies.  Having a spectrum extend to such high energies assures that GRBs are nonthermal phenomena. 

The importance of these detections is that high-energy gamma rays should not escape easily from the environment of a GRB due to the high  flux of lower-energy photons. The process of photon-photon pair production, $\gamma$ + $\gamma$ = e$^+$ + e$^-$, would effectively remove all the high-energy gamma rays, unless all the photons are moving in the same direction, i.e. beamed, with high bulk Lorentz factor (e.g. Baring 1994).  Lithwick and Sari (2001) calculate that Lorentz factors of several hundred are needed.

A second aspect of the EGRET GRB observations offers a different challenge to models of these explosions.  In addition to the prompt high-energy emission, delayed emission from some GRBs was detected by EGRET.   One example is shown in Figure 21, probably the best-known EGRET GRB detection.  In this burst, high-energy emission, including the highest-energy photon associated with a GRB at 18 GeV, was seen nearly 90 minutes after the initial burst (Hurley \etal 1994).

An example of shorter-term delayed emission was found by Gonzalez \etal (2003) by comparing BATSE data with data from the EGRET TASC calorimeter (see section 2.3).  Unlike the primary EGRET gamma-ray telescope, the TASC was an omnidirectional detector for gamma rays up to energies of nearly 200 MeV.  For GRB941017, shown in Figure 22, the GRB spectrum was seen to evolve from one dominated by the low-energy BATSE data to one dominated by the higher-energy EGRET emission over the course of about 200 seconds.  The fact that the spectrum is still rising at the upper limit of the TASC readout suggests that the delayed emission probably extended to yet higher energies. This burst was not in the field of view of the EGRET spark chamber, precluding measurements at higher energies.  The relationship of these delayed gamma-ray components of bursts may be related to afterglows seen at longer wavelengths, but such burst afterglows were not known for most of the EGRET era, so no measurements were made. 

 \begin{figure}
\centering
\includegraphics [height = 5.5 in.] {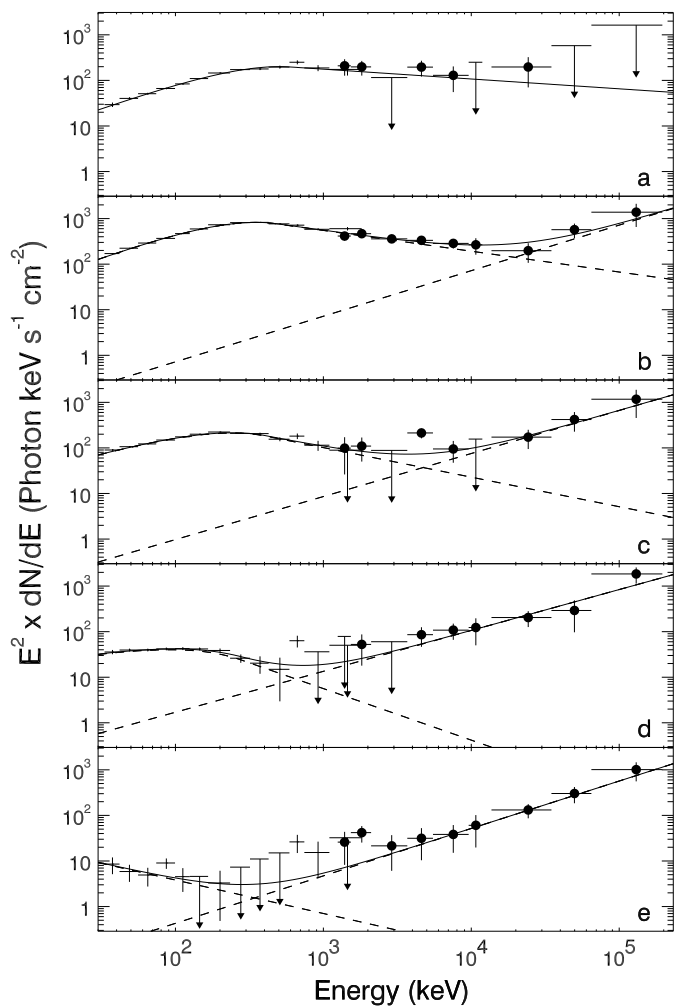}
\caption{BATSE and EGRET  spectra of GRB941017 (Gonzales \etal 2003). Crosses: BATSE; Solid circles: EGRET TASC.  The time intervals in panels a. to e. start about the time of the main burst and extend to 200 seconds later. }
\label{fig:figure22}
\end{figure}
  
\section{Local gamma-ray sources}

Although most objects energetic enough to produce gamma rays detectable by EGRET are distant, there are two sources visible within the Solar System (in addition to the Earth itself, since the Earth's atmosphere is gamma-ray bright due to cosmic ray interactions, e.g. Petry 2005).  These two local sources are the Moon and the Sun. 
 
 \subsection{The Moon}
 
 A remarkable, yet well-understood, curiosity of gamma-ray astrophysics is that in this part of the electromagnetic spectrum the Moon is brighter than the quiet Sun (when no solar flares are present).  Morris (1984) predicted the Moon to be a gamma-ray source, using a calculation extrapolated from modeling of the Earth's atmospheric gamma radiation.  The same sorts of interactions of cosmic rays with matter that produce the diffuse Galactic gamma radiation (meson production, Compton scattering, and bremsstrahlung; see section 4)  take place in the lunar surface.  Thompson \etal (1997) analyzed relevant portions of the EGRET data in a moving, Moon-centered coordinate system to verify the prediction, including evidence of the expected variation of the lunar gamma radiation with the Solar cycle of cosmic ray modulation.   Figure  23 shows the calculated and measured spectra for the Moon.  
 
 The same processes that produce the lunar radiation also operate in the atmosphere of the Sun.  All other factors being equal, the fact that the Sun and Moon subtend the same solid angle from Earth might suggest that they would be equally bright in gamma rays.  The solar magnetic field, however, excludes some of the Galactic cosmic rays from hitting the solar atmosphere, making the expected gamma-ray flux from the quiet Sun lower (Hudson, 1989; Seckel \etal 1991).  The original EGRET analysis found only an upper limit for the Sun, consistent with these calculations, although a more recent calculation and analysis of the EGRET data provides some evidence for a weak detection of the non-flare solar gamma radiation (Orlando and Strong 2008). 
 
 \begin{figure}
\centering
\includegraphics [height = 5.5 in.] {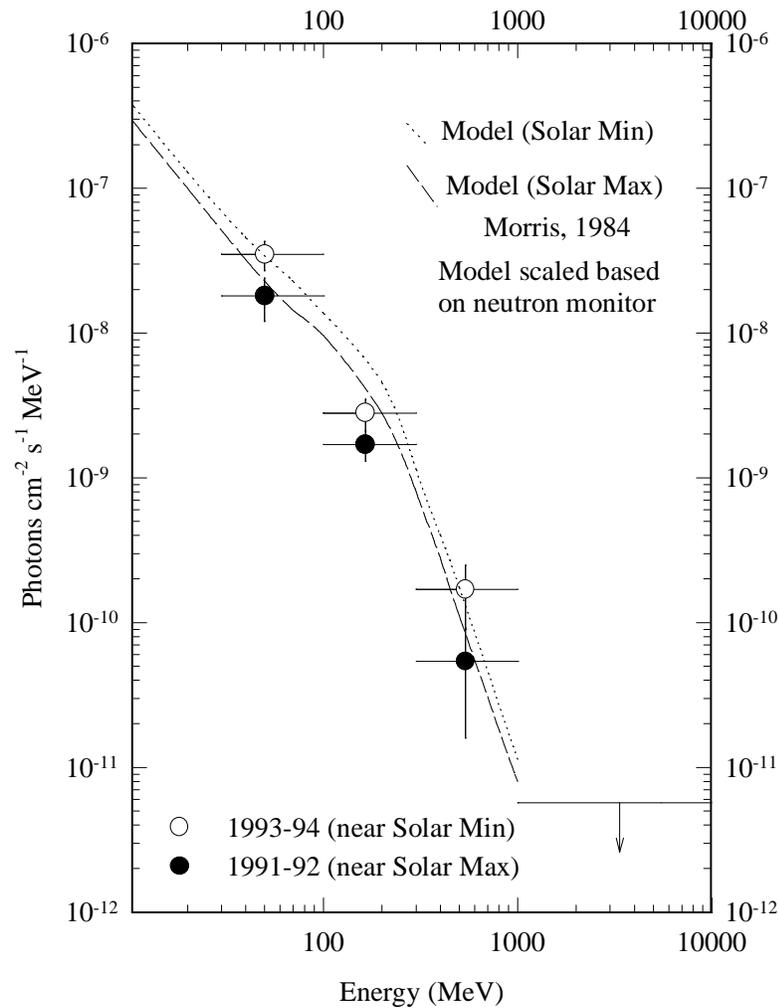}
\caption{Gamma-ray spectrum of the Moon (Thompson \etal 1996).  }
\label{fig:figure23}
\end{figure}

\subsection{Solar flares}
 
Although the quiet Sun is not a bright gamma-ray source, solar flares can accelerate particles to high enough energies to produce strong gamma-ray emission. The Compton Observatory was launched near solar maximum, and in early June of 1991 a number of solar flares were detected by all the instruments on CGRO, including EGRET (Schneid \etal 1996). On 1991 June 11 EGRET  observed gamma-ray emission  up to energies above 1 GeV for over 8 hours (Kanbach \etal 1993) following a large solar flare.  This flare represented the most energetic radiation ever seen from the Sun.  The energy spectrum confirmed evidence seen from previous missions that solar flares could accelerate protons (which collide with ambient solar material to produce neutral pions that decay into high-energy gamma rays).  That spectrum is shown in Figure 24. Modeling of the particle acceleration and interactions provided the first evidence for long-term trapping of particles in solar flare regions (Mandzhavidze and Ramaty, 1992).
 
  \begin{figure}
\centering
\includegraphics [height = 3.5 in.] {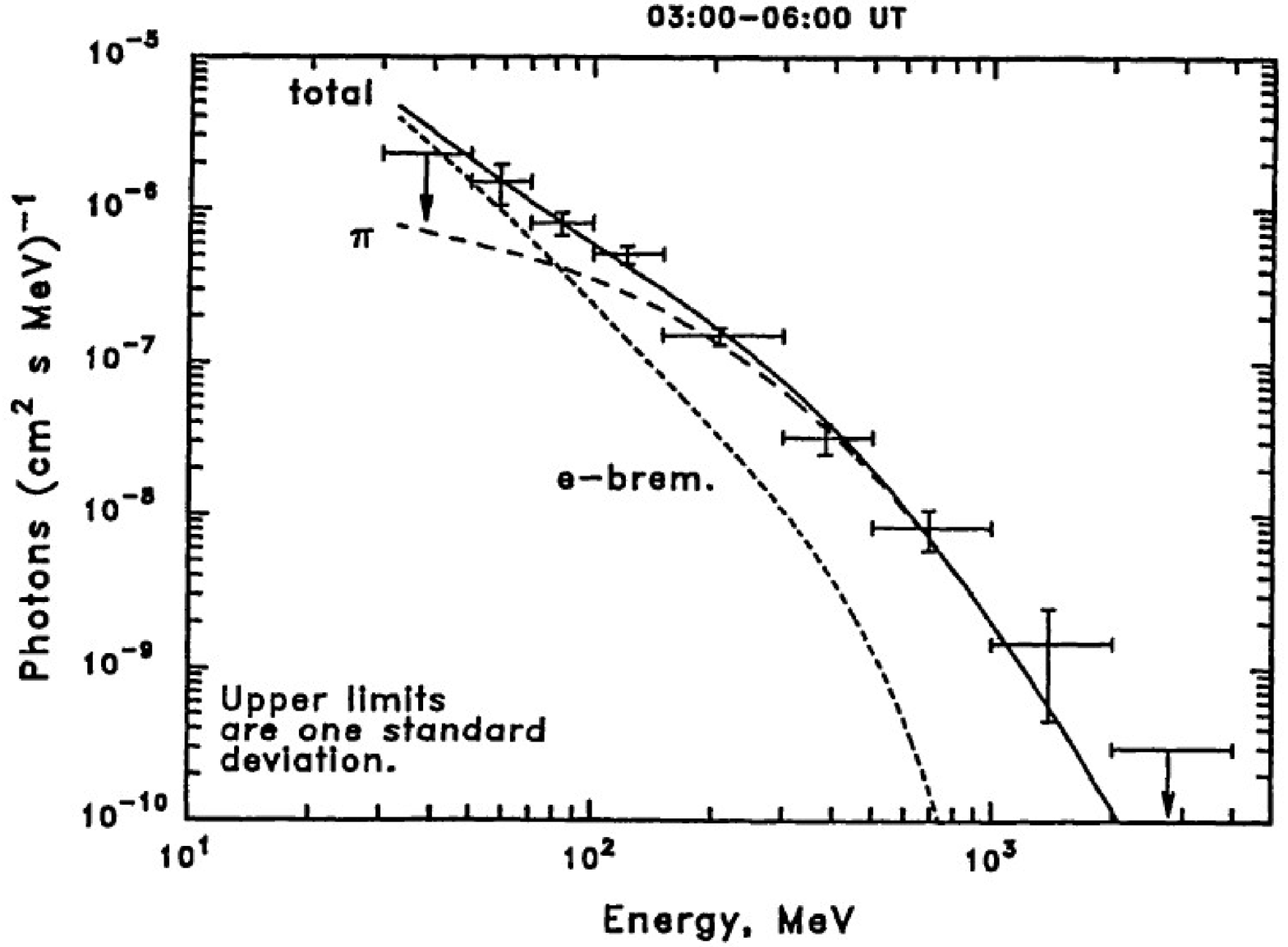}
\caption{Gamma-ray spectrum of a large solar flare (Schneid  \etal 1996.)  }
\label{fig:figure24}
\end{figure}

 \subsection{Microsecond bursts - the search for Hawking radiation from near-Earth black holes}
 
    Hawking (1974) and Page and Hawking (1976) calculated that tiny black holes that had formed in the early Universe might be detectable as flashes of high-energy gamma rays.  The process of black hole evaporation, called Hawking radiation, is negligible for solar mass black holes or larger, but those that formed with mass about 10$^{14}$ g would have lost enough mass that their final evaporation might be seen if they were close enough to the Earth.  A search of the EGRET data for multiple gamma rays occurring within the 1 microsecond live time of the spark chamber was able to set a new limit on such black holes, although not extremely restrictive (Fichtel \etal  1994 ).

\section{Summary - Open questions for AGILE and Fermi}

The Energetic Gamma Ray Experiment Telescope (EGRET) on the Compton Gamma Ray Observatory provided a dramatic new view of the high-energy Universe, including the first all-sky mapping of the sky at energies above 30 MeV.  The EGRET observations revealed a wealth of information about Galactic and extragalactic gamma radiation from both individual and diffuse sources.  Perhaps the single most striking characteristic of the EGRET gamma-ray sky is its variability, ranging from the extremely rapid flaring of gamma-ray bursts to the long-term variations seen in some sources such as blazars.  As with many successful missions, however, the questions that were answered led to new questions for future missions. With AGILE and Fermi now in operation, the time has come to seek out the solutions to some of these mysteries left behind by EGRET:

\begin{itemize}
  \item  What is the nature of the diffuse Galactic gamma radiation, and in particular the GeV excess?  Does this unexpected finding indicate some exotic new physics or a more mundane origin such as calibration?  
  \item Does the gamma radiation from the Milky Way or its surroundings contain clues to unseen forms of matter, such as cold dark gas or dark matter?  Mapping the gamma-ray emission with higher precision and comparing those measurements to information about gas derived from other types of observations is a highly promising avenue for this research.
 \item What will a larger sample of gamma-ray pulsars reveal about the location of the particle acceleration and the particle interactions processes under extreme conditions?
 \item How many more radio-quiet pulsars will be found, and what will those pulsars say about the neutron star population of our Galaxy?
 \item Which binary systems produce gamma rays, and how do those systems work?
 \item What other classes of Galactic objects have enough energy to produce gamma rays detectable by the new generation of telescopes?  \item Will including new, high-quality gamma-ray measurements of blazar spectra and time variability in multiwavelength studies provide the clues to jet properties such as composition and possibly to jet formation and collimation?
 \item What will the new gamma-ray measurements of other galaxies tell us about cosmic rays and matter densities in these systems?
 \item What other types of extragalactic objects are capable of generating detectable gamma ray fluxes?
 \item Will the new data resolve the diffuse extragalactic radiation as a collection of discrete sources, or will there be some residual diffuse emission that demands a new and possibly exotic explanation?
 \item Do most or all gamma-ray bursts have high-energy emission, and what does that radiation say about the forces at work in these explosive phenomena?
 \item Can high-energy gamma-ray measurements of solar flares shed new light on solar activity?
 \item What surprises will be found in the gamma-ray sky?
\end{itemize}

\ack
EGRET was a team effort.  My thanks go to all those who contributed, but particularly to the three Co-Principal Investigators: Carl Fichtel, Robert Hofstadter, and Klaus Pinkau.  I also greatly appreciate comments and suggestions from Bob Hartman and Olaf Reimer.

The pulsar section of this paper made use of the ATNF pulsar catalogue (Manchester \etal 2005). 

\References
\item[] Aharonian F, Buckley J, Kifune T and Sinnis G 2008  {\it Rep. Prog. Physics} {\bf 71} 1
\item[] Aharonian F \etal 2006  {\aap} {\bf 460} 743
\item[] Albert J \etal 2006 {\it Science} {\bf 312} 1771
\item[] Arons J 1996 {\aaps} {\bf 120} 49
\item[] Baring M G 1994 {\apjs} {\bf  90}  899
\item[] Berezinsky V, Hnatyk B and Vilenkin V. 2001 {\it Phys. Rev. D} {\bf 64} 043004
\item[] Bergstr\"om L, Edsj\"o J, Gustafsson M and Salati P  2006	{\it Jour. Cosmology  Astropart. Phys.} {\bf 5} 6 
\item[] Bertsch D L, Dame T M, Fichtel C E, Hunter S D, Sreekumar P, Stacy J G and Thaddeus P 1993 {\apj} {\bf  416}  587
\item[] Bertsch D L, Hartman R C, Hunter S D, Thompson D J and  Sreekumar P 2001 in {\it GAMMA2001: Gamma-Ray Astrophysics 2001}  ed.   S Ritz, N Gehrels and C R Shrader American Inst. Physics Conf. Proc, Vol. 587 (Melville, NY),  p. 706.
\item[] Bhattacharya D, Aky\"uz A, Miyagi T, Samimi J and Zych A 2003  {\aap} {\bf 404}  163
\item[] Bignami G F \etal 1975 {\it Space Sci. Instrum.} {\bf  1 }  245
\item[] Bignami G F, Caraveo P A and Lamb R C 1983 {\apj} {\bf  272 }  L9
\item[] Bignami G F, Fichtel C E, Hartman R C and Thompson D J 1979 {\apj} {\bf  323 }  649
\item[] Blandford R D and K\"onigl A 1979 {\apj} {\bf  232}  34
\item[] Blandford R D and Rees M J 1978	{\it Physica Scripta} {\bf 17} 265
\item[] Browning R, Ramsden D and Wright P J 1971 {\nat} {\bf  232}  99
\item[] Caraveo P A \etal 1980 {\aap} {\bf 91}  L3
\item[] Casanova S, Dingus B L and Zhang B 2007 {\apj} {\bf  656}  306
\item[] Cassandjian J-M and Grenier I A 2008 {\aap}  in press.
\item[] Chadwick P M, Latham I J and Nolan S J 2008  {\it J. Phys. G} {\bf 35} 1
\item[] Cheng K S and Zhang L 1998 {\apj} {\bf 498} 327
\item[] Chiang J and Mukherjee R 1998 {\apj} {\bf  496}  752
\item[] Cillis A N, Hartman R C and Bertsch D L 2004 {\apj} {\bf  601}  142
\item[] Cillis A N, Torres D F and Reimer O 2005 {\apj} {\bf  621}  139
\item[] Clark G W, Garmire G P and Kraushaar W L 1968 {\apj} {\bf  153}  L203
\item[] De Boer W  2005 {\aap} {\bf 444} 51. 
\item[] Dermer C D 2007 in {\it The First GLAST Symposium}  ed.   S Ritz, P Michelson and C Meegan American Inst. Physics Conf. Proc, Vol.921 (Melville, NY),  p. 122.
\item[] Digel S W, Hunter S D and Mukherjee R 1995 {\apj} {\bf  441} 270
\item[] Digel S W, Grenier I A, Heithausen A, Hunter S D and Thaddeus P  1996  {\apj} {\bf  463} 609
\item[] Dingus B L  2003 in {\it Gamma-Ray Burst and Afterglow Astronomy 2001: A Workshop Celebrating the First Year of the HETE Mission}, ed.   G R Ricker and R K Vanderspek. American Inst. Physics Conf. Proc, Vol. 662 (Melville, NY),  p. 240
\item[] Ensslin T A, Biermann P L, Kronberg P P and Wu X-P 1997 {\apj} {\bf  477} 560
\item[] Esposito J A \etal 1996 {\apj} {\bf  461}   820
\item[] Esposito J A \etal 1999 {\apjs} {\bf  123}  203
\item[] Fichtel C E, Hartman R C, Kniffen D A, Thompson D J, Fichtel C E, \"Ogelman H, \"Ozel M E, T\"umer T and Bignami G F 1975 {\apj} {\bf  198}  163
\item[] Fichtel C E, Bertsch D L, Hartman R C, Kniffen D A, Thompson D J, Hofstadter R, Hughes E B, Campbell-Finman L E, Pinkau K, Mayer-Hasselwander H 1983 {\it International Cosmic Ray Conference, 18th, Bangalore, India, August 22-September 3, 1983, Conference Papers}  {\bf 8}  19
\item[] Fichtel C E \etal 1994 {\apj} {\bf  434}  557
\item[] Fierro J M, Michelson P F, Nolan P L and Thompson D J 1998 {\apj} {\bf 494} 734
\item[] Fishman G J 1995 in {\it Seventeenth Texas Symposium on Relativistic Astrophysics and Cosmology} , ed.   H. B\"ohringer, G E Morfill and J E Tr\"umper. Annals of the New York Academy of Sciences, Vol. 759 (New York, NY: The New York Academy of Sciences)  p. 232
\item[] Fossati G, Maraschi L, Celotti A, Comastri A, Ghisellini G 1998 {\mnras} {\bf 299} 433
\item[] Gehrels N \etal 2000 {\nat} {\bf 404} 363
\item[] Ginzburg V  1972 {\it Nat. Phys. Sci} {\bf 239} 8
\item[] Gonzalez M M \etal 2003 {\nat} {\bf 424} 74
\item[] Grenier I A,  Casandjian J-M and Terrier R 2005 {\it Science} {\bf 307} 1292
\item[] Halpern J P \etal 2001 {\apj} {\bf  552}  L125
\item[] Halpern J P, Gotthelf E V, Mirabal N and Camilo F 2002 {\apj} {\bf  573}   L41
\item[] Harding A K, Grenier I A and Gonthier P L 2007 {\it Astrophys and Sp Science} {\bf  309}   221
\item[] Harding A K and Muslimov A 2005 {\it Astrophys and Sp Science} {\bf 297}   63
\item[] Hartman R C, Kniffen D A, Thompson D J, Fichtel C E, \"Ogelman H, T\"umer T and  \"Ozel, M. E.. 1979 {\apj} {\bf  230} 597
\item[]  Hartman R C \etal 1999 {\apjs} {\bf 123} 79
\item[]  Hartman R C \etal 1992 {\apj} {\bf 385} L1
\item[]  Hartman R C \etal 2001a {\apj} {\bf 553} 683
\item[]  Hartman R C \etal 2001b {\apj} {\bf 558} 583
\item[] Hawking S W  1974 { \nat} {\bf  248 }  30
\item[] Hermsen W \etal 1977 { \nat} {\bf  269}  494
\item[] Hewish A, Bell S J, Pilkington J D, Scott P F and Collins R A 1968 { \nat} {\bf  217}  709
\item[] Hirotani K 2008 submitted
\item[] Hudson H S 1989 in {\it Gamma Ray Observatory Science Workshop Proc.}, ed.   W N Johnson. NASA Goddard Space Flight Center, Greenbelt, MD, pp. 4-351$-$4-363,  
\item[] Hughes E B, Hofstadter R, Rolfe J, Johansson A, Bertsch D L, Cruickshank W J, Ehrmann C H, Fichtel C E, Hartman R C, Kniffen D A 1980 {\it IEEE Transactions on Nuclear Science} {\bf NS-27}  364
\item[] Hunter S D \etal  1993 {\apj} {\bf  409}  134
\item[] Hunter S D, Digel S W, de Geus E J,  and Kanbach G 1994 {\apj} {\bf  436}  216
\item[] Hunter S D \etal  1997 {\apj} {\bf  481}  205
\item[] Hurley K \etal 1994 { \nat} {\bf  372} 652 
\item[] Jorstad S G, Marscher A P, Mattox J R, Aller M F, Aller H D, Wehrle A E and Bloom S D 2001 {\apj} {\bf  556}   738
\item[] Kaaret P and Cottam J 1996 {\apj} {\bf  462}   L35
\item[] Kanbach G \etal 1993  {\aaps} {\bf  97} 349
\item[] Kanbach G \etal 1996  {\aaps} {\bf  120} 461
\item[] Kellerman K I \etal 2004  {\apj} {\bf  609} 539
\item[] Klebesadel R W, Strong I B and Olson R A 1973 {\apj} {\bf  182}  L85
\item[] Kniffen D A, Fichtel C E, Hartman R C, Thompson D J, \"Ozel M E, T\"umer T, Bignami G F and \"Ogelman H 1975 {\it Proc. 14th Internat. Cosmic Ray Conf.  (Munich)} {\bf 1} (Munich, Max-Planck-Institut f\"ur extraterrestrische Physik) 100	
\item[] Kniffen D A \etal 1993  {\apj} {\bf  411} 133
\item[] Kniffen D A \etal 1997  {\apj} {\bf  486} 126
\item[] Kramer M \etal 2003 {\mnras} {\bf 342} 1299
\item[] Kuiper L \etal 1999 {\aap} {\bf 351} 119
\item[] Kuiper L \etal 2000 {\aap} {\bf 359} 615
\item[] Kuiper L \etal 2001 {\aap} {\bf 378} 918
\item[] Kuiper L, Hermsen W, Verbunt F, Ord S, Stairs I and Lyne A 2002 {\apj} {\bf 577} 917
\item[] Kurfess J D 1996 {\aaps} {\bf 120}  5	
\item[] Leventhal M, MacCallum C J and Stang P D 1978 {\apj} {\bf  225 }  L11
\item[] Lin  Y C \etal 1992 {\apj} {\bf  401}   L61
\item[] Lithwick  Y and Sari R 2001 {\apj} {\bf  555}   540
\item[] Loeb A and Waxman E 2000 { \nat} {\bf  405} 156
\item[] Macomb D J, Gehrels N and Shrader C R 1999 {\apj} {\bf  513}   652
\item[] Mahoney W A, Ling J C, Jacobson A S and Tapphorn R M 1980 {\it Nucl. Instrum. Methods} {\bf  178}  363
\item[] Manchester R N, Hobbs G B,  Teoh A and Hobbs M 2005 {\it Astron. Jour.} {\bf 129} 1993
\item[] Mandzhavidze N and Ramaty R 1992 {\apj} {\bf 396} L111
\item[] Maraschi L \etal 1994  {\apj} {\bf  435}   L91
\item[] Mattox J R \etal 1993  {\apj} {\bf  410}   609
\item[] Mattox J R \etal 1996  {\apj} {\bf  461}   396
\item[] Mattox J R \etal 1997 {\apj} {\bf  476}   692
\item[] Mattox J R, Hartman R C and Reimer O 2001  {\apjs} {\bf  135}   155
\item[] McLaughlin M A, Mattox J R, Cordes J M and Thompson D J  1996  {\apj} {\bf  473}   763
\item[] Merck M \etal 1996  {\aaps} {\bf  120} 465
\item[] Meszaros P 2006  {\it Rep Prog Phys} {\bf  69}   2259
\item[] Miniati F 2002, {\mnras} {\bf 337}  199
\item[] Mirabal N and Halpern J P 2001 {\apj} {\bf  547}   L137
\item[] Morris D J 1984 {\it Journal of Geophysical Research}  {\bf 89} 10685
\item[] Morrison P 1958 {\NC} {\bf  7  }  858
\item[] Moskalenko I V and Strong A W 2000 {\apj} {\bf  528}  357
\item[] M\"ucke A and Pohl M 2000, {\mnras} {\bf 312} 177
\item[] Mukherjee R  2001 in {\it High-Energy Gamma-Ray Astronomy} , ed.   F A Aharonian, H J Volk, American Inst. Physics Conf. Proc, Vol. 558, (Melville, NY),  p. 324
.\item[] Mukherjee R and Chiang J 1999 {\it APh} {\bf  11} 213
\item[] Mukherjee R, Halpern J, Mirabal N and Gotthelf E V  2002 {\apj} {\bf  574}  693
\item[] Mukherjee R  \etal 1995 {\apj} {\bf  441}  L61
\item[] Mukherjee R  \etal 1997 {\apj} {\bf  490}  116
\item[] Mukherjee R  \etal 1999, {\apj} {\bf  527}  132
\item[] Nandikoktkur G \etal 2007 {\apj} {\bf  657} 706
\item[] Nolan P L \etal 2003 {\apj} {\bf  597} 615
\item[] Orlando E and Strong A W 2008 {\aap} {\bf 480} 847
\item[] \"Ozel M E and Thompson D J 1996 {\apj} {\bf  643}  105
\item[] Page  D N  and Hawking S 1976 {\apj} {\bf  206}  1
\item[] Paredes J M,  Mart' J,   Rib\'o M,  and Massi M 2000 {\it Science} {\bf 288} 2340
\item[] Pavlidou V and Fields B D 2002  {\apj} {\bf  575} L5
\item[] Petry D  2005 in {\it High-Energy Gamma-Ray Astronomy} , ed.   F. A. Aharonian, H. J. Volk., D. Horns, American Inst. Physics Conf. Proc, Vol. 745, (Melville, NY),  p. 709
\item[] Punch M \etal 1992 { \nat} {\bf 358} 477
\item[] Reimer O 2001 in: {\it The Nature of the Unidentified Galactic Gamma-Ray Sources}, ed. A. Carrimi\~nana, O Reimer, D J Thompson, Astrophysics and Space Science Library Vol. 267, Kluwer (Dordrecht), p.17
\item[] Reimer O and Thompson D J 2001 in {\it Proc. 27th International Cosmic Ray Conference, Hamburg, Germany}, Copernicus Gesellschaft, p. 2566. 
\item[] Reimer O \etal 2001 {\mnras} {\bf 324} 772	
\item[] Reimer O, Pohl M, Sreekumar P and Mattox J R 2003  {\apj} {\bf  558} 155
\item[] Riegler G R, Ling J C, Mahoney W A, Wheaton W A, Willett J B, Jacobson A S and Prince T A 1981 {\apj} {\bf 248}  L13
\item[] Roberts M S E, Romani R W and Kawai N 2001 {\apjs} {\bf  133}  451 
\item[] Romani R W  1996, {\apj} {\bf 470} 469
\item[] Romero G E \etal 1999 {\aap} {\bf  348} 868R
\item[] Ruderman M A and Sutherland P G 1975 {\apj} {\bf 196}  51
\item[] Schneid E J \etal 1996 {\aaps} {\bf  120}  299
\item[] Sch\"onfelder V (ed.) 2001 {\it The Universe in Gamma Rays}   Berlin, Springer
 \item[] Sch\"onfelder V, Bennett K, Bloemen H, Diehl R, Hermsen W, Lichti G, McConnell M, Ryan J, Strong A and Winkler C 1996 {\aaps} {\bf 120}  13 
\item[] Sguera V, Bassani L, Malizia A, Dean A J, Landi R and Stephens J B 2005 {\aap} {\bf 430} 107
\item[] Sikora  M, Begelman M C and  Rees M J 1994 {\apj} {\bf  421}  153
\item[] Sowards-Emmerd D, Romani R W, Michelson P F and Ulvestad J S 2004 {\apj} {\bf  609}  564
\item[] Sreekumar P  \etal  1992 {\apj} {\bf  400}  L67
\item[] Sreekumar P  \etal  1993 {\it Phys. Rev. Lett.} {\bf  70}  127
\item[] Sreekumar P  \etal  1994 {\apj} {\bf  426}  105
\item[] Sreekumar P  \etal  1998 {\apj} {\bf  494}  523
\item[] Sreekumar P,  Bertsch D L,  Hartman R C, Nolan P L and Thompson D J 1999 {\it Astropart. Phys} {\bf 11} 221 
\item[] Steckel D, Stanev T  and Gaisser T K 1991  {\apj} {\bf  382}  652
\item[] Stecker F and Salamon M 1996  {\apj} {\bf  464}  600
\item[] Stecker F W, Hunter S D and Kniffen D A  2008 {\it Astropart. Phys} {\bf 29}  25
\item[] Strong A W, Moskalenko I V and Reimer O 2000 {\apj} {\bf 537 }  763
\item[] Strong A W, Moskalenko I V and Reimer O 2004a {\apj} {\bf 613 }  956
\item[] Strong A W, Moskalenko I V and Reimer O 2004b {\apj} {\bf 613 }  962
\item[] Sturner S  J and Dermer C D 1995   {\aap} {\bf 293} L17
\item[] Sturrock P 1971 {\apj} {\bf 164 } 529
\item[] Swanenburg B N, Hermsen W, Bennett K, Bignami G F, Caraveo P, Kanbach G, Mayer-Hasselwander H A, Masnou J L, Paul J A, and Sacco B. 1978 { \nat} {\bf  275}  298
\item[] Swanenburg B N \etal 1981 {\apj} {\bf 243 }  L69
\item[] Tavani M   \etal  1997 {\apj} {\bf  479}  L109
\item[] Tavani M,  Kniffen D, Mattox J R, Paredes J M and Foster R S  1998 {\apj} {\bf  497}  L89
\item[] Teshima M 2008  {\it The Astronomer's Telegram} 1491
\item[] Thompson D J  in {\it Cosmic Gamma-Ray Sources}, ed. K S Cheng and G E Romero (Kluwer, Dordrecht Boston London 2004) p 149
\item[] Thompson D J, Fichtel C E, Kniffen D A and \"Ogelman H B 1975 {\apj} {\bf  200}  L79
\item[] Thompson D J \etal 1993  {\apjs} {\bf 86}  629
\item[] Thompson D J \etal 1995  {\apjs} {\bf 101}  259
\item[] Thompson D J, Bertsch D L, Morris D J and Mukherjee R 1997 {\it Journal of Geophysical Research}  {\bf 102} 14735
\item[] Thompson D J, Digel S W, Nolan P L and Reimer O 2002
in: {\it Neutron Stars in Supernova Remnants}, ed. P O Slane, B M Gaensler, (Boston), ASP Conference Series 271,  p.65
\item[] Thompson D J,  Bertsch D L and Hartman R C  2001 in {\it GAMMA2001: Gamma-Ray Astrophysics 2001}, ed.   S Ritz, N Gehrels and C R Shrader. American Inst. Physics Conf. Proc, Vol. 587 (Melville, NY),  p. 668
.\item[] Thompson T A, Quataert E and Waxman E 2007 {\apj} {\bf 654}  219
\item[] Tornikoski M \etal 1999   {\it Astron. J.} {\bf 118  }  1161
\item[] Torres D,  Romero G E, Dame T M, Combi J A and Butt Y M   {\it Phys. Reports} {\bf 382}  303
\item[] Ullio P, Bergstr\"om L, Edsj\"o J and Lacey C 2002 {\it Phys. Rev. D} {\bf 66} 123502 
\item[] Urry C M and  Padovani  P 1995 {\it PASP} {\bf 107} 803 
\item[] Vestrand T  \etal 1997 {\apj} {\bf  483}  L49
\item[] Von Montigny C \etal 1995a {\aap} {\bf  299} 680
\item[] Von Montigny C \etal 1995b {\apj} {\bf 440} 525
\item[] Wagner S J \etal 1995 {\apj} {\bf  454}  L97
\item[] Wallace P M, Griffis N J, Bertsch D L, Hartman R C, Thompson D J, Kniffen D A and Bloom S D 2000 {\apj} {\bf  540}  184
\item[] Weekes T C \etal 1989 {\apj} {\bf  373}  289
\item[] Weekes T C 2003  {\it Very High Energy Gamma-Ray Astronomy} (Bristol: Institute of Physics Publishing)
\item[] Wehrle A E \etal 1998 {\apj} {\bf  497}  178
\item[] Yadigoraglu I A and Romani R W 1997 {\apj} {\bf  476}  347

\endrefs

\end{document}